\documentclass[preprint2]{aastex631}
\usepackage{amsmath}
\usepackage{ulem}
\usepackage{hyperref}
\setcitestyle{notesep={ }}
\usepackage{bm}
\usepackage{tensor}
\usepackage{tikz}
\usetikzlibrary{positioning}
\usetikzlibrary{patterns}
\usetikzlibrary{decorations.pathmorphing}
\usetikzlibrary{arrows.meta}
\usetikzlibrary{decorations.markings}
\usetikzlibrary{shapes.geometric}

\graphicspath{{./}{figures/}}

\definecolor{ao(english)}{rgb}{0.0, 0.5, 0.0}

\newcommand{\aperture}{\emph{Aperture}}

\begin{document}

\title{Introducing APERTURE: A GPU-based General Relativistic Particle-in-Cell Simulation Framework}

\correspondingauthor{Alexander Y. Chen}
\email{cyuran@wustl.edu}

\author[0000-0002-4738-1168]{Alexander Y. Chen}
\affil{Physics Department and McDonnell Center for the Space Sciences, Washington University in St. Louis; MO, 63130, USA}

\author[0009-0008-1825-6043]{Martin Luepker}
\affil{Physics Department and McDonnell Center for the Space Sciences, Washington University in St. Louis; MO, 63130, USA}

\author[0000-0002-0108-4774]{Yajie Yuan}
\affil{Physics Department and McDonnell Center for the Space Sciences, Washington University in St. Louis; MO, 63130, USA}

\begin{abstract}
    Low-luminosity Active Galactic Nuclei (AGN) are believed to be surrounded by
    a collisionless, highly magnetized accretion flow. As a result,
    Particle-in-Cell simulations are the best tools to study the immediate
    vicinity of the event horizons of these supermassive black holes. We present
    a GPU-based general relativistic particle-in-cell (GRPIC) code framework
    called \aperture{}. \aperture{} is developed in C++, with compute kernels written
    in CUDA and HIP to take advantage of the massive acceleration modern GPUs
    enable. The code is organized in a fully modular way, allowing easy extensions
    to new physics problems. In this paper, we describe in detail the particle
    pusher, field solver, and charge-conserving current deposition algorithms
    employed in \aperture{}, and present test cases to validate their correctness.
    Then, we apply the code to study spark gaps and plasma injection in black hole magnetospheres. We find that the apparent location and time-evolution of the gap depend on the observer. Our results reconcile the previous conflicting findings from 1D and 2D simulations in the literature.
\end{abstract}

\keywords{
 black holes ---
 numerical methods ---
 plasma astrophysics
}

\section{Introduction} \label{sec:intro}
Black holes power a wide range of fascinating astrophysical phenomena. For example, Active Galactic Nuclei (AGN) are fueled by accretion of gas onto the supermassive black holes residing at the centers of galaxies; some of these black holes can launch powerful, relativistic jets that reach far beyond the host galaxy and shape the environment around the galaxy. Stellar mass black holes within our own galaxy can also accrete gas from their companions and exhibit similar behaviors on a smaller scale. Our understanding of the physics behind these phenomena has come a long way, thanks to unprecedented coverage of multiwavelength observations, and the progress in theoretical modeling, especially large scale numerical simulations that became possible with modern computational power.

General relativistic magnetohydrodynamics (GRMHD) simulations have been widely used to study the accretion of gas onto the black hole and the launching of jets, providing invaluable insights. For instance, they have shown that black holes can indeed launch jets through the Blandford-Znajek process~\citep{1977MNRAS.179..433B}---a rapidly spinning black hole threaded by a strong magnetic field can have its rotational energy extracted electromagnetically~\citep[e.g.,][]{2011MNRAS.418L..79T,2012MNRAS.423.3083M}. However, GRMHD has its limitations, as it is built on the assumption that the plasma is collisional and locally in thermal equilibrium. In low luminosity AGN, due to the low accretion rate, the gas density can be so low in the vicinity of the black hole that the collisional mean free path becomes comparable or larger than the dynamic length scale (e.g., the gravitational radius of the black hole, $r_g=GM/c^2$) \citep{2014ARA&A..52..529Y}. The behavior of such a collisionless plasma can not be properly described by MHD. In particular, the plasma can develop pressure anisotropy or even nonthermal distributions~\citep[e.g.,][]{2014PhRvL.112t5003K}. The rate of dissipation in ideal MHD may also be different from the collisionless limit~\citep{2009PhPl...16k2102B,2014ApJ...783L..21S,2014PhRvL.113o5005G,2016ApJ...816L...8W}. We can only recover the rich physics by going to a more fundamental approach, e.g.\ starting from kinetic descriptions.

In recent years, general relativistic particle-in-cell (GRPIC) simulations
became possible due to a combination of advances in computational power and
numerical algorithms.
One dimensional (1D) GRPIC simulations were first used by
\citet{2018A&A...616A.184L} and \citet{2020ApJ...895..121C} to study the spark
gaps in black hole magnetospheres. In particular, \citet{2020ApJ...895..121C}
found that along an arbitrary field line in a monopolar magnetosphere,
macroscopic vacuum gaps can open quasiperiodically, producing bursts of
$e^{\pm}$ pairs and high energy radiation. The first 2D GRPIC simulation was
carried out by \citet{2019PhRvL.122c5101P}, who studied jet launching when the
black hole is immersed in an asymptotically uniform magnetic field. It was found
that with sufficient plasma supply, the structure of the magnetosphere is
similar to the force-free configuration, such that all the field lines that go
through the ergosphere eventually enter the event horizon and rotate with the
black hole. When plasma supply gets low, the current in the jet column is
conducted by a charge separated flow and particles with negative energy at
infinity contribute significantly to black hole rotational energy extraction
through a variant of the Penrose process.

Subsequently, \citet{2020PhRvL.124n5101C} used 2D GRPIC simulations to study the
pair discharge process around a spinning black hole with a monopolar magnetic
field. They found highly time-dependent spark gaps open near the inner light
surface which inject pair plasma into the magnetosphere.
\citet{2021A&A...650A.163C, 2022PhRvL.129t5101C} also carried out 2D and 3D GRPIC
simulations of more realistic black hole magnetosphere configurations where a
large scale equatorial current sheet exists and a conducting disk mimics an
accretion disk to transport magnetic flux toward the black hole. They obtained
synthetic gamma-ray light curves and radio images from these kinetic
simulations. 
\citet{2021PhRvL.127e5101B} used GRPIC simulations to study the balding of a Kerr black hole initially endowed with a dipole field, and found that the field opens into a split-monopole and reconnects in a plasmoid-unstable current sheet; as a result, the field on the black hole decays exponentially with time. 
\citet{2022A&A...663A.169E,2023A&A...677A..67E} performed 2D and 3D
GRPIC simulations of spinning black holes magnetically connected to a Keplerian
disk. They showed that the field lines connecting the black hole and the disk
can transfer a significant amount of angular momentum and energy between the
black hole and the disk; above the loop top where opposite open field lines
meet, a current sheet develops and magnetic reconnection takes place, which may
be a heat source for the so-called corona. Recently, the spark gaps in black
hole magnetospheres have been further studied using 1D
\citep{2020ApJ...902...80K,2024ApJ...964...78K} and 2D
\citep{2023MNRAS.526.2709N} GRPIC simulations, confirming the time dependent
nature of the gaps, and the necessity of pair injection between the two light
surfaces. Furthermore, large scale 2D GRPIC simulations of collisionless,
spherical accretion onto a spinning black hole have been carried out by
\citet{2023PhRvL.130k5201G}, who demonstrated that collisionless magnetic
reconnection makes a difference in the flow dynamics compared to GRMHD
simulations. They also showed that in the collisionless plasma, there can be
significant departures from thermal equilibrium, including pressure anisotropy
that excites kinetic-scale instabilities, and a large field-aligned heat flux
near the horizon.

These first-principles simulations have proven to be very fruitful. The
capability of GRPIC simulations is opening up a new window for us to peer into
the rich physics of plasmas around black holes. There are a lot more questions
to be answered, especially the ones concerning particle acceleration and
emission near the black hole, jet launching, its mass loading, and the
interaction between the jet and the disk/wind. Kinetic physics plays an important
role in these problems, and GRPIC simulations are going to provide us new
insights into solving these long-standing puzzles.

In this paper, we present a GPU-based GRPIC code framework called \aperture{}.
\aperture{} is a recursive acronym that stands for ``\textbf{A}perture simulates
\textbf{P}articles, \textbf{E}lectrodynamics, and \textbf{R}adiative
\textbf{T}ransfer at \textbf{U}ltra-\textbf{R}elativistic \textbf{E}nergies''.
The first version of \aperture{} was developed as a part of the author Chen's
doctoral thesis~\citep{2017PhDT.......278C}, and was first used to study the
magnetospheres of pulsars~\citep{2014ApJ...795L..22C}. It was the first
  GPU-based PIC code applied to the field of high energy plasma astrophysics.
Over the years, the codebase has been through several significant rewrites in
attempt to find a stable, modular infrastructure that is extensible and easy to
maintain. This paper presents the fourth and most recent major version, which
includes MPI parallelization, GPU acceleration, radiative transfer, and
capabilities for different coordinate systems such as Cartesian, cylindrical,
and spherical coordinates. The most recent addition is general relativity in 2D
axisymmetry, which further expands the realm of applicability of this code.

This paper is dedicated to describing the GRPIC implementation utilized in
\aperture{} and a physics application, with only brief discussions about its
architectural design and technical aspects. The paper is organized as follows.
Section~\ref{sec:algorithms} introduces the major algorithms behind \aperture{}
that make GR calculations possible. Section~\ref{sec:code-validation} presents a
series of test cases that validate each major area of the code, ending with an
integrated test in the form of the plasma-filled Wald solution.
Section~\ref{sec:application} uses the code to conduct a numerical experiment
related to plasma production in black hole magnetospheres and compares our
finding with previous work. Finally, Section~\ref{sec:discussions} discusses the
prospects of this code in the community as well as the potential scientific
problems that it enables us to study.

\section{The GRPIC Algorithm} \label{sec:algorithms}

In \aperture{}, we employ the standard $3+1$ split of the
spacetime~\citep{2004MNRAS.350..427K}. The spacetime metric is written as:
\begin{equation}
    \label{eq:metric}
    ds^{2} = (\beta^{2} - \alpha^{2})\,dt^{2} + 2\beta_{i}\,dx^{i}dt + \gamma_{ij}\,dx^{i}dx^{j},
\end{equation}
where $\alpha$ is called the ``lapse function'' and $\bm{\beta}$ is called the
``shift vector''. We follow the convention of using Greek letters such as $\mu$, $\nu$
to denote 4-dimensional spacetime indices, and alphabets such as $i$, $j$ to denote
3-dimensional spatial indices. The spatial part of the full spacetime metric is
$\gamma_{ij}$, which coincides with the metric of the 3-space slices given by
constant global time $t$.

In order to avoid singularities at the event horizon, we use the Kerr-Schild
coordinates $(t, r, \theta, \phi)$, similar to~\citet{2019PhRvL.122c5101P}. The
Kerr-Schild coordinate observers are on a series of infalling trajectories, and
the shift vector $\bm{\beta}$ has only radial component $\beta^{r}$. A summary
of the metric and relevant quantities is given in
Appendix~\ref{app:kerr-schild}. In order to avoid ambiguity, we generally
refrain from using vector notation such as $\bm{B}$ in equations, but always
prefer to explicitly state whether we are referring to contravariant vectors
$B^{i}$ or covariant vectors $B_{i}$, where the spatial metric $\gamma_{ij}$ and its
inverse $\gamma^{ij}$ are used to raise or lower the indices. The Einstein
summation notation is assumed in the rest of the paper, unless explicitly stated
otherwise.

The standard PIC loop consists of three main components: particle update,
current deposit, and field update~\citep[see e.g.][]{1991ppcs.book.....B}.
Modern PIC codes often introduce extra physics such as radiation, collision, and
pair production, but they are often added on top of the classic PIC loop. In
order to implement a minimal working GRPIC code, at least the three basic
algorithms need to be modified to include the GR corrections. In the rest of
this section, we describe how the three algorithms are modified in the framework
of GR.

\subsection{Field Solver}
\label{sec:field-solver}

We use the contravariant vector fields $D^{i}$ and $B^{i}$ as dynamic variables
for the electric and magnetic field, which are defined from the Maxwell tensor
$F_{\mu\nu}$ as:
\begin{equation}
    \label{eq:D-B}
    D^{i} = \alpha F^{0i},\quad B^{i} = \alpha\tensor[^*]{F}{^{i0}},
\end{equation}
where the dual of the Maxwell tensor is defined by:
\begin{equation}
    \label{eq:Fdual}
    \tensor[^{*}]{F}{_{\mu\nu}} = e_{\mu\nu\alpha\beta}F^{\alpha\beta}
\end{equation}
where $e_{\mu\nu\alpha\beta} = \sqrt{-g}\epsilon_{\mu\nu\alpha\beta}$ is the Levi-Civita pseudo-tensor of spacetime and $\epsilon_{\mu\nu\alpha\beta}$ is the 4-dimensional completely anti-symmetric Levi-Civita symbol.

The Maxwell field equations in curved spacetime are:
\begin{equation}
    \label{eq:D-B-evolution}
    \begin{split}
      \nabla_i D^i &= \frac{1}{\sqrt{\gamma}}\partial_i \left(\sqrt{\gamma} D^i\right) = \rho, \\
      \nabla_i B^i &= \frac{1}{\sqrt{\gamma}}\partial_i \left(\sqrt{\gamma} B^i\right) = 0, \\
      \partial_t D^i &= e^{ijk}\partial_j H_k - J^i, \\
      \partial_t B^{i} &= -e^{ijk}\partial_{j}E_{k}.
    \end{split}
\end{equation}
Here we employ the Heaviside-Lorentz units which absorbs factors of $4\pi$. We
also choose a natural unit system and set $c = 1$ in the equations. The symbol
$e_{ijk}$ is the 3-dimensional Levi-Civita pseudo-tensor,
$e_{ijk} = \sqrt{\gamma}\epsilon_{ijk}$ and
$e^{ijk} = \epsilon_{ijk}/\sqrt{\gamma}$. The 3-current $J^{i}$ is related to
the 4-current $I^{\mu}$ by $J^{i} = \alpha I^{i}$, and the charge density $\rho$ is $\rho = \alpha I^0$. The first two equations are simply constraints that we need to enforce on initial conditions, while the two time-dependent equations are what we evolve in a PIC simulation. The auxiliary fields $\bm{E}$ and
$\bm{H}$ are defined as:
\begin{equation}
    \label{eq:E-H}
    \begin{split}
    E_{i} &= F_{i0} = \frac{1}{2}\alpha e_{ijk}\tensor[^{*}]{F}{^{jk}}, \\
    H_{i} &= \tensor[^{*}]{F}{_{0i}} = \frac{1}{2}\alpha e_{ijk}F^{jk}.
    \end{split}
\end{equation}
One can write these in terms of the dynamic fields:
\begin{equation}
    \label{eq:E-H-B-D}
    \begin{split}
      E_i &= \alpha \gamma_{ij}D^j + e_{ijk}\beta^jB^k, \\
      H_i &= \alpha \gamma_{ij}B^j - e_{ijk}\beta^jD^k.
    \end{split}
\end{equation}

Equations~\eqref{eq:D-B-evolution} resemble the vacuum Maxwell equations, except
that their right hand sides involve the same fields being updated due to Equations~\eqref{eq:E-H-B-D}, which makes the equations numerically more challenging.

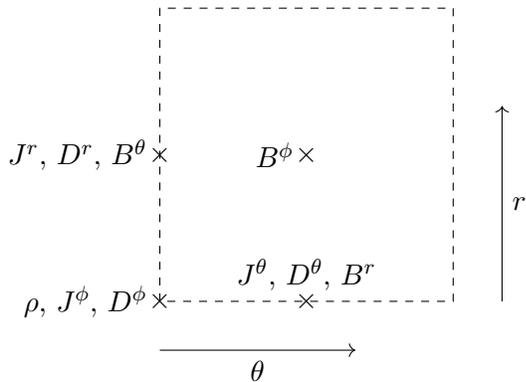
\begin{figure}[h]
  \centering
  \begin{tikzpicture}[label distance=-2.5mm]
    \begin{scope}[scale=1.3]
      \draw[->] (3.5, 0) -- node[right] {$r$} (3.5, 2);
      \draw[->] (0, -0.5) -- node[below] {$\theta$} (2, -0.5);
      \draw[dashed] (0,0) rectangle (3,3);
      \node [label=left:$B^\phi$] (center) at (1.5,1.5) {$\times$};
      \node [label=left:$J^{r}\text{, }D^r\text{, }B^{\theta}$] (right) at (0,1.5) {$\times$};
      \node [label=above:$J^\theta\text{, }D^\theta\text{, }B^{r}$] (top) at (1.5,0) {$\times$};
      \node [label=left:$\rho\text{, }J^{\phi}\text{, }D^{\phi}$] (top right) at (0,0) {$\times$};
    \end{scope}
  \end{tikzpicture}
  \caption{Staggered Yee cell for the 2D grid used in \aperture{}. The magnetic field $\bm{B}$ is defined at the center of cell faces, while the electric field $\bm{D}$ and current density $\bm{J}$ is defined on the middle of cell edges. The electric charge density $\rho$ is defined on cell vertices.}
  \label{fig:Yee}
\end{figure}

To solve the field equations numerically, we adopt a staggered Yee
grid~\citep{1966ITAP...14..302Y}, where the $\bm{B}$ field is face-centered
while the $\bm{D}$ field is edge-centered (see Figure~\ref{fig:Yee}). Instead of
directly working with Equations~\eqref{eq:D-B-evolution}, we integrate the
equations over a cell face $\Sigma_{i}$, and assume that the dynamic fields are constant over
the small face area. Then, the time evolution equations become the integral form of the generalized Maxwell equations (no Einstein summation over $i$):
\begin{equation}
    \label{eq:integral-form}
    \begin{split}
      \partial_tD^iA^{D}_i &= \sum_{\mathrm{sides}\ j} H_{j} \Delta x^j - J^iA^{D}_i, \\
      \partial_tB^iA^{B}_i &= -\sum_{\mathrm{sides}\ j} E_{j} \Delta x^{j},
    \end{split}
\end{equation}
where we have defined the area element
\begin{equation}
    \label{eq:area-element}
    A_{i} = \int_{\Sigma_{i}}\frac{1}{2}e_{ijk}dx^{j}\wedge dx^{k} = \int_{\Sigma_i} \sqrt{\gamma}dx^{j}dx^{k}.
\end{equation}
Note that due to the staggered definition of $\bm{D}$ and $\bm{B}$, their area elements $A_{i}$ are offset by half a grid cell, therefore we included a superscript in Equation~\eqref{eq:integral-form} to distinguish them. The integral in Equation~\eqref{eq:area-element} is evaluated for every cell face at the beginning of the simulation using a simple 10--point Gauss quadrature. Since the integrand is smooth over each cell, and the integration volume is usually small, the Gauss quadrature can achieve very high precision for the purpose of the simulation.

We solve Equations~\eqref{eq:integral-form} using the discrete forms that are
given in Appendix~\ref{app:field-eqn-discrete}. In the code, we first evaluate
the auxiliary fields $H_{i}$ and $E_{i}$, then use them to update the dynamic
fields $D^{i}$ and $B^{i}$. Since these equations are implicit, we use a
predictor-corrector scheme to iteratively compute $D^{n+1,i}$ and $B^{n+1,i}$
from $D^{n,i}$ and $B^{n,i}$, where $n$ labels time step indices. We first carry out the ``predictor'' step:
\begin{equation}
    \label{eq:fld-predictor}
    \begin{split}
      \tilde{D}^{n+1,i} &= D^{n,i} + (C_{H}^{i}(\bm{D}^{n}, \bm{B}^{n}) - J^{n+1/2,i})\Delta t, \\
      \tilde{B}^{n+1,i} &= B^{n,i} + C_{E}^{i}(\bm{D}^{n}, \bm{B}^{n})\Delta t,
    \end{split}
\end{equation}
where $C_{H}$ and $C_{E}$ are the circulations of the auxiliary fields $H$ and
$E$, as defined in Appendix~\ref{app:field-eqn-discrete}. Then, we apply the ``corrector'' step with an interpolation between steps $n$ and $n+1$:
\begin{equation}
    \label{eq:fld-corrector}
    \begin{split}
      D^{n+1,i} &= D^{n,i} + \frac{\Delta t}{2}\left(\alpha C_{H}^{i}(\bm{D}^{n}, \bm{B}^{n}) + \phantom{\frac{1}{2}} \right. \\
      &\left. \beta C_{H}^{i}(\tilde{\bm{D}}^{n+1}, \tilde{\bm{B}}^{n+1})\right) - J^{n+1/2,i}\Delta t, \\
      B^{n+1,i} &= B^{n,i} + \frac{\Delta t}{2}\left(\alpha C_{E}^{i}(\bm{D}^{n}, \bm{B}^{n}) + \phantom{\frac{1}{2}} \right. \\
      &\left. \beta C_{E}^{i}(\tilde{\bm{D}}^{n+1}, \tilde{\bm{B}}^{n+1})\right).
    \end{split}
\end{equation}
We have inserted a pair of semi-implicit interpolation coefficients $\alpha$ and $\beta$ such that $\alpha + \beta = 1$, similar to~\citet{2013PhPl...20f2904H}. When $\alpha = \beta = 0.5$, this scheme is essentially the Heun's method and is second order accurate. When $0.5 < \beta \leq 1$, the scheme can damp out high frequency waves due to its decentered implicit nature, while retaining approximately second order accuracy as long as $\beta$ is close to $0.5$. To improve accuracy, we iterate the corrector step $N$ times. In production, we typically choose $\beta \approx 0.53$ and $N \sim 7$.

The staggered nature of the Yee grid ensures the vanishing of $\nabla\cdot \bm{B} \equiv \nabla_i B^i$ to machine precision, regardless of how $E_{i}$ is computed. Similarly, as long as the current deposition scheme is charge-conserving, $\nabla\cdot \bm{D} \equiv \nabla_i D^i$ is also guaranteed to be equal to the charge density. We will describe our current deposition algorithm in Section~\ref{sec:current-deposition} and verify its validity in Section~\ref{sec:plasma-wald}.

The boundary conditions for 2D GR simulations are as follows. The staggering of the field components are chosen such that only $D^{r}$, $B^{\theta}$, and $D^{\phi}$ are defined on the coordinate axis. As a result, no special conditions need to be applied to $B^{\phi}$, $D^{\theta}$, and $B^{r}$ at $\theta$ boundaries. For the components on the axis, we set $D^{\phi} = B^{\theta} = 0$ due to symmetry requirements. For $D^{r}$ on the axis at $\theta = 0$, we update it using the following equation:
\begin{equation}
    \label{eq:Dr-axis}
    D^{n+1,r}_{i0} = D^{n,r}_{i0} + \Delta t\left(\frac{H_{\phi,i0}}{A^{D}_{r,i0}}\right) - J^{r}_{i0},
\end{equation}
where $i$ labels the cell index in $r$, subscript 0 means the cell on the axis, and $A^{D}_{r,i0}$ only integrates from $\theta = 0$ to $\theta = \Delta \theta/2$. The current $J^{r}$ is computed at the boundary cell with correct reflection imposed so that no particle crosses the axis. This way charge conservation is guaranteed numerically even at the axis. We update $D^{r}$ at the $\theta = \pi$ axis using a similar scheme, replacing $H_{\phi,i0}$ by $-H_{\phi,i,N_{\theta}-1}$ due to the staggering.

The inner $r$ boundary is chosen to be slightly within the event horizon of the black hole. We find that setting $r_\mathrm{in} = 1$, or equivalently $\log r_\mathrm{in} = 0$, is usually sufficient and easy to implement. At this boundary we apply the usual zero derivative condition, setting the values of the field components in the guard cells to be the same as the last physical cell. This allows the horizon to behave reasonably well, and no unphysical perturbations propagate outwards from within the horizon. At the outer $r$ boundary, we use a damping layer that gradually reduces all field components to zero. We find it to allow particles to peacefully flow out of the simulation domain, causing negligible amount of wave reflection.

\subsection{Particle Pusher}
\label{sec:particle-pusher}

We use the contravariant 3-coordinate $x^{i}$ and the covariant momentum vector
$u_{i}$ as our dynamic variables for charged particles in the simulation. The
benefit of such a choice is that angular momentum $u_{\phi}$ is manifestly
conserved when there is no electromagnetic field, hence improving the numerical
stability of the algorithm. The particle equations of motion under the $3+1$
split of the spacetime are~\citep[see e.g.][]{2018ApJS..237....6B}:
\begin{equation}
    \label{eq:particle-eom}
    \begin{split}
      \frac{dx^{i}}{dt} &= \gamma^{ij}\frac{u_{j}}{u^{0}} - \beta^{i}, \\
      \frac{du_{i}}{dt} &= -\alpha u^{0}\partial_{i}\alpha + u_{k}\partial_{i}\beta^{k} - \frac{u_{j}u_{k}}{2u^{0}}\partial_{i}\gamma^{jk} + \frac{F_{i}}{m},
    \end{split}
\end{equation}
where $F_{i}$ is the covariant vector for the Lorentz force: 
\begin{equation}
    \label{eq:Lorentz}
    F_{i} = q\left(\alpha\gamma_{ij}D^{j} + e_{ijk}\gamma^{jl}\frac{u_{l}}{u^{0}}B^{k}\right).
\end{equation}
The time component $u^{0}$ is computed from the momentum $u_{i}$ as $u^{0} = \sqrt{\epsilon + \gamma^{ij}u_{i}u_{j}}/\alpha$ where $\epsilon = 1$ or $0$ depending on whether the particle is massive (electron, positron, or ion) or massless (photon). Photons are treated as a separate particle species that do not experience the Lorentz force $F_{i}$, but their motion is still described by the geodesic equation~\eqref{eq:particle-eom}.

Similar to~\citet{2019PhRvL.122c5101P}, we employ a Strang split to solve the
particle's geodesic motion and Lorentz force update separately. We use an
iterative second-order predictor-corrector scheme that is similar to the field
solver to perform the geodesic update, and adopt a generalized Boris algorithm to
apply the Lorentz force. At every time step, we first update the particle
momentum $u_{i}$ by a half step using the generalized Boris algorithm, then carry
out the geodesic update by a full time step. Finally, we update $u_{i}$ by
another half step using the new particle location and new fields. For photons,
the Boris step is skipped.

The geodesic update proceeds as follows. Using $X^{i}$ and $U_{i}$ to denote the right hand sides of Equations~\eqref{eq:particle-eom} (excluding the Lorentz force term $F_{i}$), we first carry out the ``predictor'' step:
\begin{equation}
    \label{eq:geodesic-predictor}
    \begin{split}
      \tilde{x}^{n+1, i} &= x^{n,i} + X^{i}(\bm{x}^{n}, \bm{u}^{n})\Delta t, \\
      \tilde{u}^{n+1}_{i} &= u^{n}_{i} + U_{i}(\bm{x}^{n}, \bm{u}^{n})\Delta t.
    \end{split}
\end{equation}
Then, we apply the ``corrector'' step with a trapezoidal rule:
\begin{equation}
    \label{eq:geodesic-corrector}
    \begin{split}
      x^{n+1, i} &= x^{n,i} + \frac{\Delta t}{2}\left(X^{i}(\tilde{\bm{x}}^{n+1}, \tilde{\bm{u}}^{n+1}) + X^{i}(\bm{x}^{n}, \bm{u}^{n})\right), \\
      u^{n+1}_{i} &= u^{n}_{i} + \frac{\Delta t}{2}\left(U_{i}(\tilde{\bm{x}}^{n+1}, \tilde{\bm{u}}^{n+1}) + U_{i}(\bm{x}^{n}, \bm{u}^{n})\right).
    \end{split}
\end{equation}
Again, the corrector step can be applied iteratively for a number of times. In
practice, we find that a few iterations already yields very good results (see
Section~\ref{sec:trajectories}). The default number of iterations is 3 in the
production \aperture{} code.

Due to the predictor-corrector method used, our particle position $x^{i}$ and
momentum $u_{i}$ are both defined at integer time steps $n$, in contrast to typical
PIC codes that use a leapfrog update scheme. This conveniently lines up with the half-step Boris push, which uses either the field values at time step $n$ or $n+1$. The momentum update at step $n$ looks like the following:
\begin{equation}
    \label{eq:boris-step}
    u_{i}^{\prime n} = u_{i}^{n} + \frac{q}{m}\Delta t\left(\alpha\gamma_{ij}D^{n, j} + e_{ijk}\gamma^{jl}\frac{\bar{u}_{l}^{n}}{\bar{u}^{0}}B^{n, k}\right),
\end{equation}
where similar to the regular Boris push, a mid-step momentum $\bar{u}^{n}$ and
``Lorentz factor'' $\bar{u}^{0}$ are needed. It is customary to define
$\bar{u}^{n} = (u^{n} + u^{\prime n})/2$, which makes Equation~\eqref{eq:boris-step} an implicit equation for $u^{\prime n}$ that needs to be solved. In order to perform the update
explicitly without iteration, we use the same algorithm
as~\citet{boris_relativistic_1970}, defining:
\begin{equation}
    \label{eq:boris-plus-minus}
    \begin{split}
      u_i^- &= u_i^{n} + \frac{q}{m}\frac{\alpha \Delta t}{2}\gamma_{ij}D^{n, j},\\
      u_i^+ &= u_i^{\prime n} - \frac{q}{m}\frac{\alpha \Delta t}{2}\gamma_{ij}D^{n, j},\\
      t^i &= \frac{q B^{n, i}}{2 m \bar{u}^0}\Delta t,
    \end{split}
\end{equation}
where $\bar{u}^{0}$ is computed using $u_{i}^{-}$:
\begin{equation}
    \label{eq:u0bar}
    \bar{u}^{0} = \frac{1}{\alpha}\sqrt{1 + \gamma^{ij}u_{i}^{-}u_{j}^{-}}.
\end{equation}
Then, it can be shown that the following scheme, together with Equations~\eqref{eq:boris-plus-minus}, correctly solves Equation~\eqref{eq:boris-step}:
\begin{equation}
    \label{eq:boris-scheme}
    \begin{split}
      \tilde{u}_i &= u_i^- + e_{ijk}\gamma^{jl}u_l^-t^k,\\
      s^i &= \frac{2t^i}{1+t^kt_k},\\
      u_i^+ &= u_i^-+e_{ijk}\gamma^{jl}\tilde{u}_l s^k.
    \end{split}
\end{equation}
This is the straightforward generalization of the Boris pusher to curved spacetime. Since the same metric coefficients are used throughout the process, the cost of this algorithm is not too different from transforming to the local FIDO frame, as is done by~\citet{2019PhRvL.122c5101P} and~\citet{2020PhRvL.124n5101C}.

\begin{figure}[h]
    \centering
    \begin{tikzpicture}[
        node distance=3cm and 1cm,
        arrow/.style={-Stealth, thick},
        fieldnode/.style={draw, circle, minimum width=0.1cm},
        particlenode/.style={draw, circle, minimum width=0.1cm},
        ]

        \foreach \t/\label in {0/n-1, 1/n, 2/n+1} {
          \node[draw, rectangle] (t\t) at (\t*3.2,0) {$t_{\label}$};
        }

        \foreach \t/\label in {0/n-1, 1/n, 2/n+1} {
          \node[particlenode] (P\t) at (\t*3.2,-1.3) {};
          \node[above left] at (P\t.north west) {$x^i, u_{i}$};
        }

        \foreach \t in {0,1,2} {
          \draw[arrow, red] (P\t) to [out=50,in=110,loop,looseness=12.8] node[midway,right=0.1cm] {$u_{i}'$} (P\t);
        }
        \foreach \t/\label in {0/n-1, 1/n, 2/n+1} {
          \node[fieldnode] (F\t) at (\t*3.2,-3.1) {};
          \node[below=0.3cm] at (F\t) {$D^{i},B^{i},\rho$};
        }

        \foreach \t/\label in {0/n-\frac{1}{2}, 1/n+\frac{1}{2}} {
          \node (J\t) at (\t*3.2+1.6,-2.65) {$J^{i}_{\label}$};
        }

        \foreach \t in {0,1} {
          \pgfmathtruncatemacro{\nextT}{\t+1}
          \draw[arrow] (F\t) -- (F\nextT);
          \draw[arrow, blue, out=330,in=210] (P\t) to node[midway,above] (M\t) {} (P\nextT);
          \draw[arrow, blue] (M\t) -- (J\t);
        }

    \end{tikzpicture}
    \vspace{-0.2in}
    \caption{The GRPIC time stepping scheme of \aperture{}. The field quantities $D^{i}$, $B^{i}$, and the charge density $\rho$ are defined at integer time steps $n$. The particle positions $x^{i}$ and momenta $u_{i}$ are also defined at integer time steps. A semi-implicit iterative scheme is used to update the particle position and momentum to the next integer time step (blue arrows), then an additional Boris push is performed to update $u_{i}$ to $u_{i}'$ using the local electromagnetic field (red arrows). The current density $J^{i}$ is computed from the difference between $x^{i,n+1}$ and $x^{i,n}$, which naturally lands at half integer time steps. We then directly use $J^{i}$ at half time steps to evolve the field quantities.}
    \label{fig:stepping}
\end{figure}
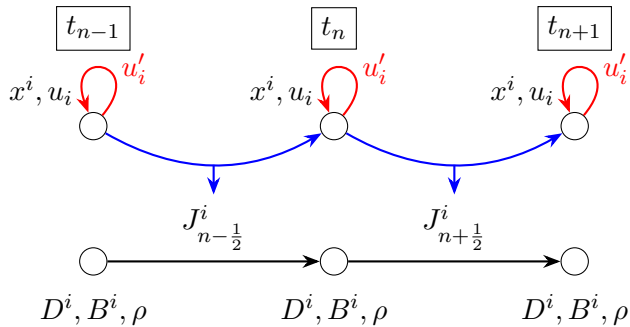

Note that we follow~\citet{2019PhRvL.122c5101P} in splitting the Boris momentum update and the geodesic update into a sequence of a half-step Boris, followed by a full step geodesic, then another half-step Boris at the new particle location. A choice can be made at the second half-step Boris update whether to use the updated field values at time step $n+1$, or to use the field from previous time step $n$. Since this second half-step push uses the particle positions at time step $n+1$ (see Figure~\ref{fig:stepping} for illustration), it can be argued that it's self-consistent to use the field values at step $n+1$ as well. In this case, this half-step Boris push becomes exactly equivalent to the initial Boris push of the next time step, and thus can be bundled together into a full step Boris push. We have experimented with both choices and have not found any discernable difference in global PIC simulation results. Since bundling the two adjacent Boris half-steps saves one field interpolation, it is the convention we adopt in \aperture{}. An additional Vay pusher is also implemented in the code where we carry out a transformation to the FIDO frame and perform the Vay momentum update there~\citep{2008PhPl...15e6701V}, although the benchmarks in this paper are exclusively using the generalized Boris algorithm.

\subsection{Current Deposition}
\label{sec:current-deposition}

In order to achieve charge-conserving current deposition, \aperture{} uses a
modified version of the Esirkepov current deposition
algorithm~\citep{2001CoPhC.135..144E}. The goal is to determine the electric
current $J^{i}$ in a way that is consistent with the 4-dimensional continuity
equation $\nabla_{\mu}I^{\mu} = 0$, which in 3+1 formalism reads:
\begin{equation}
    \label{eq:charge-continuity}
    \partial_{t}\rho + \nabla_{i}J^{i} = 0,
\end{equation}
where $\rho = \alpha I^{0}$ and $J^{i} = \alpha I^{i}$. Similar to the field solver, we use the integral form of this equation in the code:
\begin{equation}
    \label{eq:J-integral-eqn}
    \begin{split}
      \frac{\Delta \rho_{i, j}\Delta V_{i, j}}{\Delta t} &= -(J^{r}_{i,j}A^{D}_{r,i,j} - J^{r}_{i-1,j}A^{D}_{r,i-1,j}) \\
      &\quad -(J^{\theta}_{i, j}A^{D}_{\theta, i,j} - J^{\theta}_{i, j-1}A^{D}_{\theta, i, j-1}),
    \end{split}
\end{equation}
where subscripts $i$, $j$ labels the cell indices in the $r$ and $\theta$
directions, respectively. Since the electric current $J^{i}$ is defined at the
same location as the electric field $D^{i}$ (see Figure~\ref{fig:Yee}), its area
elements are the same as $A^{D}_{i}$.

Equation~\eqref{eq:J-integral-eqn} is formally equivalent to the Cartesian
version if we identify the combinations $J^{r} A_{r}$ and
$J^{\theta} A_{\theta}$ as the Cartesian electric current, therefore we can use
the same density decomposition proposed by~\citet{2001CoPhC.135..144E} to split $\Delta \rho$ into $\Delta \rho_{r}$ and $\Delta \rho_{\theta}$ for each particle:
\begin{equation}
    \label{eq:esirkepov-decomp}
    \begin{split}
      \Delta \rho_{r,ij} &= \frac{q}{2}\left[S_{ij}(r^{n+1},\theta^{n+1}) - S_{ij}(r^{n}, \theta^{n+1})\right. \\
      &\quad + \left.S_{ij}(r^{n+1}, \theta^{n}) - S_{ij}(r^{n}, \theta^{n})\right] / \Delta V_{ij}, \\
      \Delta \rho_{\theta, ij} &= \frac{q}{2}\left[S_{ij}(r^{n+1},\theta^{n+1}) - S_{ij}(r^{n+1}, \theta^{n})\right. \\
      &\quad + \left.S_{ij}(r^{n}, \theta^{n+1}) - S_{ij}(r^{n}, \theta^{n})\right] / \Delta V_{ij},
    \end{split}
\end{equation}
where $S_{ij}$ is the particle shape function, evaluated at the grid cell $(i, j)$, and $q$ is the charge of the given particle. In a current deposit loop, we evaluate $\Delta\rho_{r}$ and $\Delta\rho_{\theta}$ for each individual particle, associating $J^{r}$ with $\Delta\rho_{r}$ and $J^{\theta}$ with $\Delta\rho_{\theta}$. Then, we use a simple prefix sum to compute $J^{r}A_{r}^{D}$ and $J^{\theta}A^{D}_{\theta}$ caused by the movement of this particle. Since the particle shape functions have finite support, this prefix sum is carried out locally for each particle over several cells. Finally, after finishing the current deposition loop, we divide the results by $A_{r}^{D}$ and $A_{\theta}^{D}$ to get the actual current densities $J^{r}$ and $J^{\theta}$ respectively. The shape function $S_{ij}$ used in \aperture{} is the first order cloud-in-cell interpolation by default, but a third order spline is also available for use through a configuration option.

The azimuthal current $J^{\phi}$ does not need to undergo a prefix sum since it is not part of the continuity equation. We simply compute $J^{\phi}$ using the Esirkepov average:
\begin{equation}
    \label{eq:j-phi-deposit}
    \begin{split}
    J^{\phi}_{ij} &= -q\frac{\phi^{n+1} - \phi^{n}}{\Delta t}\left[ \frac{1}{3}S_{ij}(r^{n+1}, \theta^{n+1})\right. \\
      &\quad + \frac{1}{6}S_{ij}(r^{n+1}, \theta^{n}) \\
                    &\quad +\left.\frac{1}{6}S_{ij}(r^{n},\theta^{n+1}) + \frac{1}{3}S_{ij}(r^{n}, \theta^{n})\right].
    \end{split}
\end{equation}

There are several benefits to the algorithm described above. First, it is
manifestly charge-conserving, as the same terms $J^{r}A_{r}^{D}$ and
$J^{\theta}A_{\theta}^{D}$ are used to update the electric fields $D^{r}$ and
$D^{\theta}$ according to Equation~\eqref{eq:integral-form}. Second, it does not
involve any branching conditions, as opposed to the standard Buneman
scheme~\citep{1992CoPhC..69..306V} or its
derivatives~\citep[e.g.][]{2003CoPhC.156...73U}. This is very beneficial on the GPU
architecture where branch divergence within a warp can be costly~\citep[see e.g.][]{cuda}.
Finally, this current deposition algorithm only depends on the initial and final
positions of each particle over the time step, but not on its velocity. As a
result, we do not need to compute the contravariant particle velocity $v^{i}$ from
$u_{i}$ during current deposition, nor do we need to interpolate it to the half time step $n+1/2$. We will present tests for the extent of charge conservation in Section~\ref{sec:plasma-wald}.

\subsection{Aperture}
\label{sec:aperture}

\aperture{} is written in C++, heavily utilizing the modern features introduced
up to the \verb|c++17| standard, especially template metaprogramming. The
organization of the code is inspired by the Entity, Component, System (ECS)
paradigm of modern video game engines. Each major module (e.g.\ field solver) in
the \aperture{} code is a \emph{system}, and each data structure (e.g.\ particle
array) is a \emph{data component}. A problem setup is defined by an ordered list
of systems, a collection of data components, together with some initial
condition for these data components. At every time step during the PIC loop, the
\verb|update()| method of each system is called to act on one or multiple data
components, e.g.\ updating the positions and momenta of each particle in the
simulation. Such a paradigm allows for easy extension and dependency management,
as each system defines the data structures it acts on, and registers them with
the main simulation framework. A given simulation setup is uniquely defined by
the list (and the order) of systems and initial conditions for all relevant data
components, making it trivial to create new setups without affecting existing
ones. The flexibility of the framework allow completely new systems to be implemented that are not necessarily limited to PIC, but may also include Force-free electrodynamics or potentially other physics algorithms.

At the \emph{system} level, \aperture{} uses a common C++ approach called
\emph{policy-based design}~\citep{10.5555/377789}, which allows the developer to
alter the behavior of a class using a variety of flexible components
called \emph{policies}. For example, the \verb|ptc_updater| class that handles
particle update in \aperture{} takes 3 policies as its template parameters. The
first policy specifies whether the code execution is on CPU or GPU; the second
one specifies the coordinate system; and the third policy can be used to
implement extra physics such as gravitational forces. To extend the particle
updater to include GR, only a new coordinate policy needs to be developed, which
reuses as much code as possible and saves a tremendous amount of work.

\begin{deluxetable*}{cccccccc}[t]
\centerwidetable
\tablecaption{Parameters and initial conditions for test particle trajectories}
\tablehead{\colhead{Name} & \colhead{\hspace{.75cm}Spin}\hspace{.5cm} & \colhead{\hspace{.75cm}$B_z$}\hspace{.5cm} & \colhead{\hspace{.75cm}$u_r$}\hspace{.5cm} & \colhead{\hspace{.75cm}$u_\theta$}\hspace{.5cm} & \colhead{\hspace{.75cm}$u_\phi$}\hspace{.5cm} & \colhead{ \hspace{.75cm}$r_i$}\hspace{.5cm} & \colhead{\hspace{.75cm}$\theta_i$}\hspace{.5cm}}
\startdata
(2,0,1) & 0.0 & 0.0 & 0.136491 & 0 & 3.9 & 16.109371 & $\pi/2$ \\
(3,3,1) & 0.995 & 0.0 & 0.201211 & 0 & 2 & 10.021533 & $\pi/2$ \\
(3,3,2) & 0.995 & 0.0 & 0.487625 & 0 & 1.82 & 4.376281 & $\pi/2$ \\
(4,3,1) & 0.995 & 0.0 & 0.204751 & 0 & 2 & 9.851412 & $\pi/2$ \\
RKA2 & 0.7 & 2.0 & 0.921532 & 0 & -1.93 & 4.2 & $\pi/2-0.1$ \\
RKA3 & 0.9 & 2.0 & 0.638604 & 0 & 1.565 & 4 & $\pi/2-0.2$ \\
RKA8 & 0.9 & 1.0 & 1.15912 & 0.497 & 0.365 & 3 & $\pi/2$
\enddata
\tablecomments{All parameters are given in code units. For all orbits, the initial $\phi$ angle is $\phi_i=0$. The values and naming conventions come from~\citet{2008PhRvD..77j3005L} and~\citet{2019ApJS..240...40B} respectively.}
\label{tab:paths}
\end{deluxetable*}

Implementation of code execution on GPU is done mainly through a wrapper which can execute a \verb|for| loop in a parallel fashion, with the content of the loop provided by a \verb|lambda| function. This is the simplest way to allow for portability between CPU and GPU, as well as between CUDA and ROCm platforms. In the case of CUDA kernels, the launch parameters (e.g.\ block sizes and grid sizes) are automatically determined by maximizing the kernel residency.
The full source code of \aperture{} is freely available on GitHub and under
continuous
development\footnote{\href{https://github.com/fizban007/Aperture4}{https://github.com/fizban007/Aperture4}}.
It is released under the GPLv3 license, which permits free use, modification,
and redistribution of the code under the same license.

\section{Code Validation}
\label{sec:code-validation}

\begin{figure*}[t]
\centering
\begin{minipage}{0.53\textwidth}
\includegraphics[width=\textwidth]{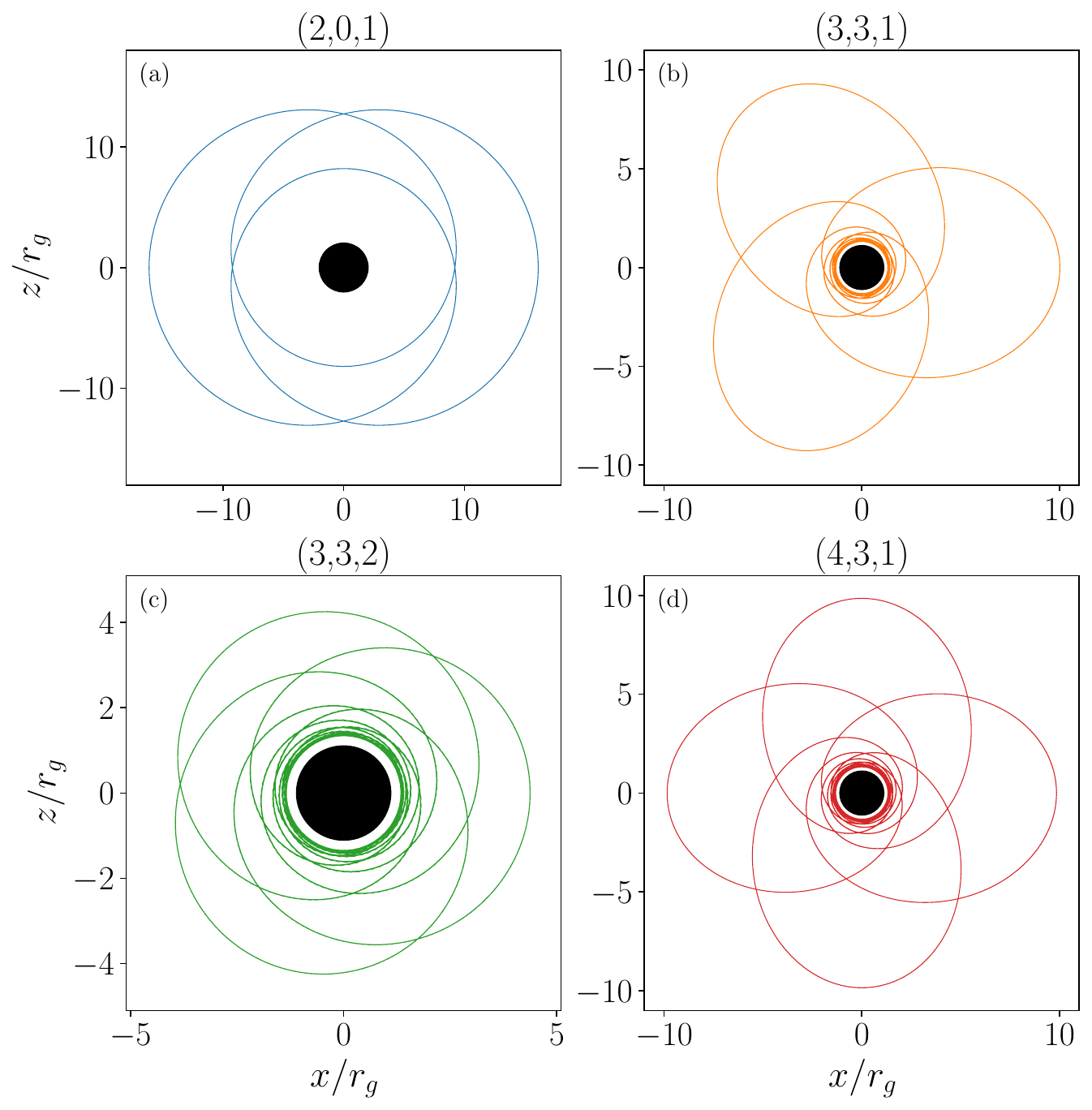}
\end{minipage}
\begin{minipage}{0.45\textwidth}
\includegraphics[width=\textwidth]{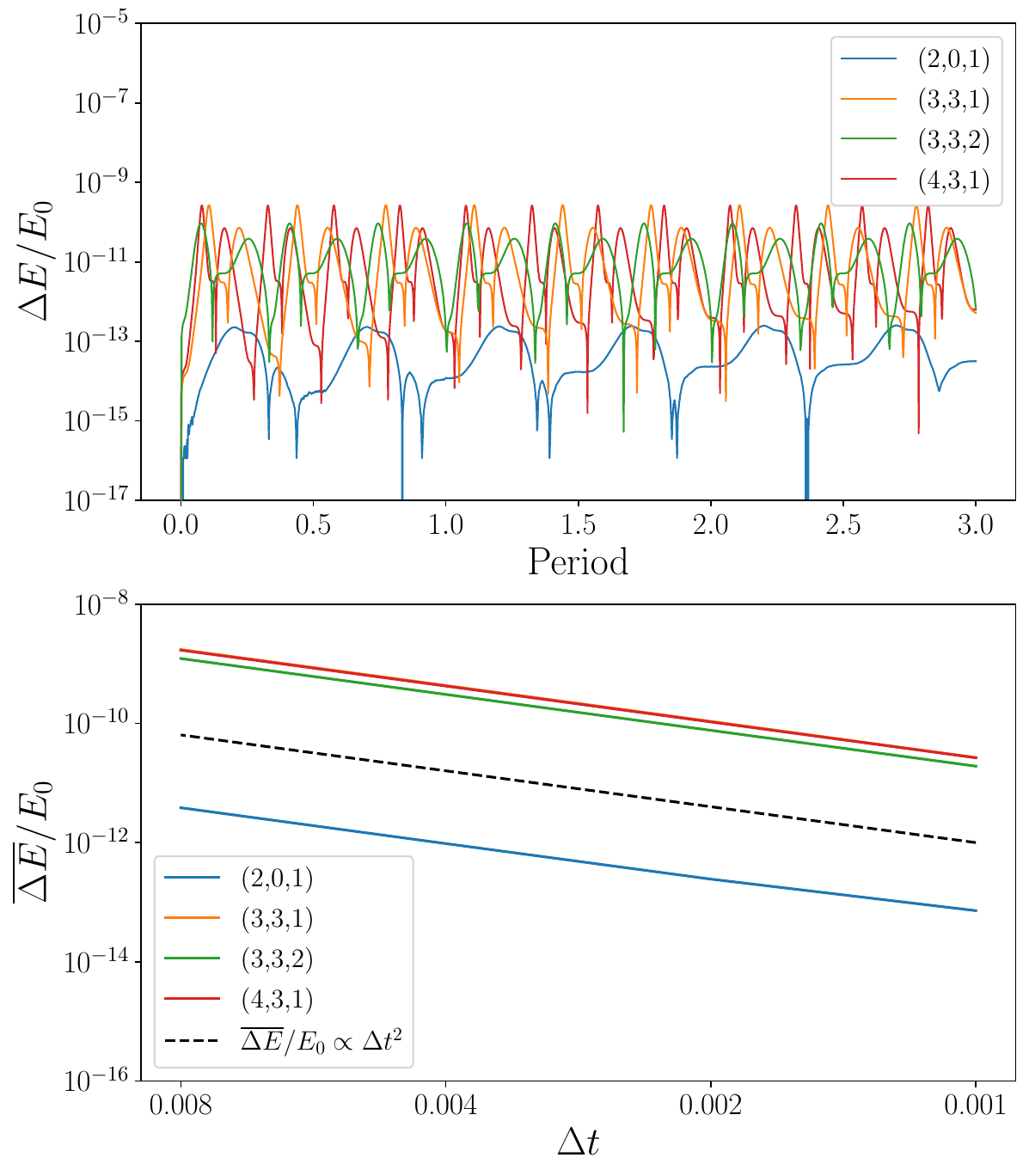}
\end{minipage}
\caption{Neutral periodic equatorial particle orbits around black holes. \emph{Left:} Path (a) is on a non-spinning black hole $a = 0$ and the other three are on spinning black holes $a = 0.995$. Each path is integrated for three periods. \emph{Top Right:} Deviation of energy from its original value, $\Delta E = |E - E_{0}|$, vs time for each path. \emph{Bottom Right:} Convergence of the energy deviation over a period with respect to varying time step size $\Delta t$. The relative error decreases as $\Delta t^{2}$, which is expected for a second order scheme. }
\label{fig:orbits}
\end{figure*}

\begin{figure*}[hbt!]
\centering
\includegraphics[width=\textwidth]{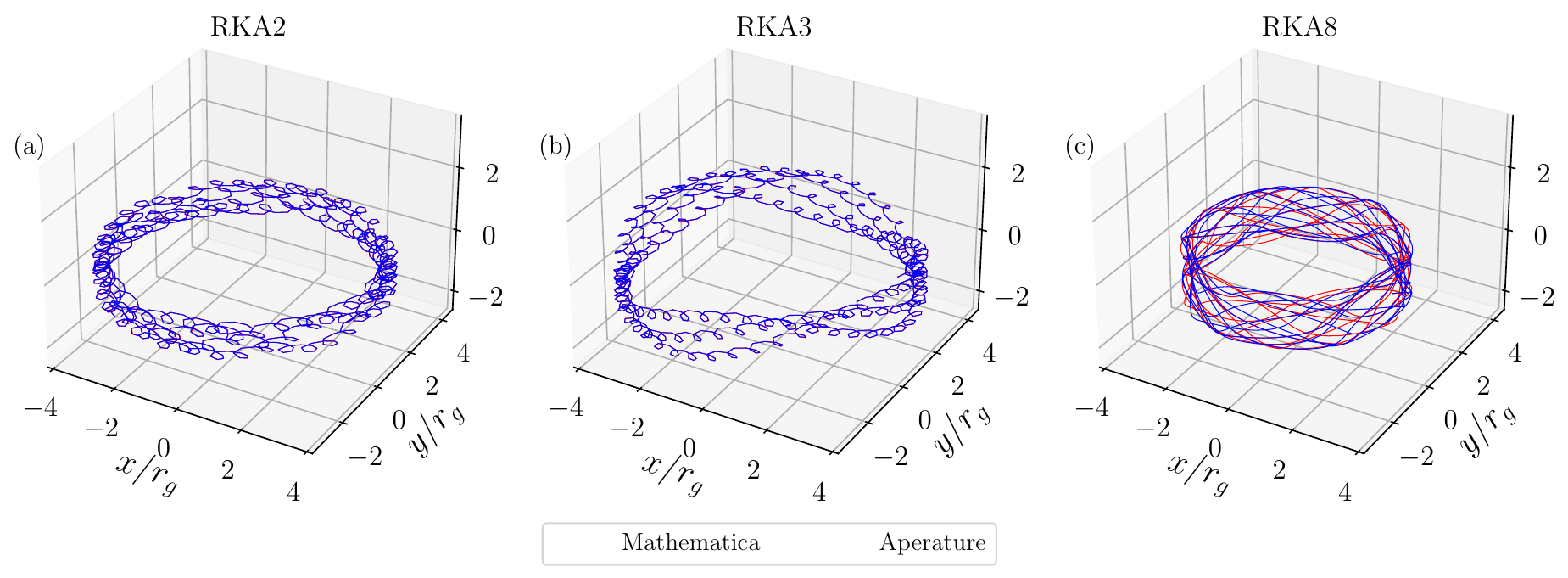}
\caption{Comparison between charged trajectories evaluated by \aperture{} and Mathematica. The \aperture{} trajectories are evaluated with grid resolution of $N_{r} = N_{\theta} = 1024$ and $\Delta t = 0.001r_{g}/c$. The first two cases agree extremely well, while the third test case quickly accumulates a finite amount of phase error, leading to deviation of the trajectories.}
\label{fig:charged_orbits}
\end{figure*}

In this section, we present a series of test cases for validating the algorithms described in Section~\ref{sec:algorithms}. First, we integrate the trajectories of neutral and charged particles in \aperture{} and compare them with known high-precision solutions. Then, we test the electromagnetic field solver in vacuum. Finally, we combine both components and reproduce the plasma-filled Wald solution of the black hole magnetosphere.

\subsection{Test Particle Trajectories}
\label{sec:trajectories}

We integrate the paths of particles around black holes to test \aperture{}'s
particle mover. First, we consider only the geodesic solver, in the absence of any electromagnetic field. We use equatorial periodic geodesic paths for
  massive particles described by~\citet{2008PhRvD..77j3005L}. Among
  the zoo of periodic orbits, we have selected 4 particular paths in both Schwarzschild and Kerr metrics.
The orbits are labeled by a triplet of integers $(z, w, v)$ that describe topological properties of the paths. Among the 4 orbits, (a) is for a Schwarzschild black hole, while the other 3 are for a Kerr black hole with spin $a = 0.995$. The initial conditions are listed in Table~\ref{tab:paths}.
These paths are a useful tool for testing \aperture{}'s particle
mover since the periodicity gives a test of stability over time and we can
examine the gravitational forces independent of the electromagnetic ones.

The left panels of Figure~\ref{fig:orbits} show the 4 neutral particle geodesics, which are produced in \aperture{} and integrated for approximately 3 full periods. The right panels show the evolution of the conserved energy $E = -u_{0}$ over time, and the energy deviation as we decrease the time step size $\Delta t$. Instead of directly plotting $E$, we plot the deviation from the initial energy, $\Delta E = |E - E_{0}|$, which allows us to quantify the numerical error involved in the calculation. As can be seen in the graph, the geodesic solver in \aperture{} is capable of conserving the particle energy over the entire duration of integration for all 4 particle trajectories. Since the energy fluctuates over time, we plot $\overline{\Delta E}$ which is $\Delta E$ averaged over a period in the convergence plot. We find that all cases are consistent with $\overline{\Delta E}\propto \Delta t^{2}$, which is expected for a second order scheme.

Next, we evaluate the paths of charged particles in a vacuum Wald field, similar to the approach by ~\citet{2019ApJS..240...40B}. These paths test the combined effects of gravitational and electromagnetic forces which is important for realistic plasma simulations. We have selected 3 orbits, labeled as RKA2, RKA3, and RKA8 following~\citet{2019ApJS..240...40B}. These are regular orbits that are less sensitive to cumulative numerical errors, as opposed to the chaotic orbits where small deviations grow exponentially. The initial conditions for these trajectories are listed in Table~\ref{tab:paths}.
We compare the solution obtained in \aperture{} with a high-accuracy explicit Runge-Kutta solver. The reference solution is obtained with the \verb+NDSolve+ routine in Mathematica, using
the ``ExplicitRungeKutta'' method with a high enough \verb+AccuracyGoal+ $(90)$ and \verb+MaxSteps+ $(100000)$ to ensure convergence. Figure~\ref{fig:charged_orbits} shows a comparison between \aperture{} and Mathematica evaluated on
the three test charged particle orbits.
The first two paths show very good agreement, while the third exhibits a growing phase shift, indicating the orbit is likely already close to chaotic regime in the bifurcation diagram.

\begin{figure}[h]
    \centering
    \includegraphics[width=0.48\textwidth]
{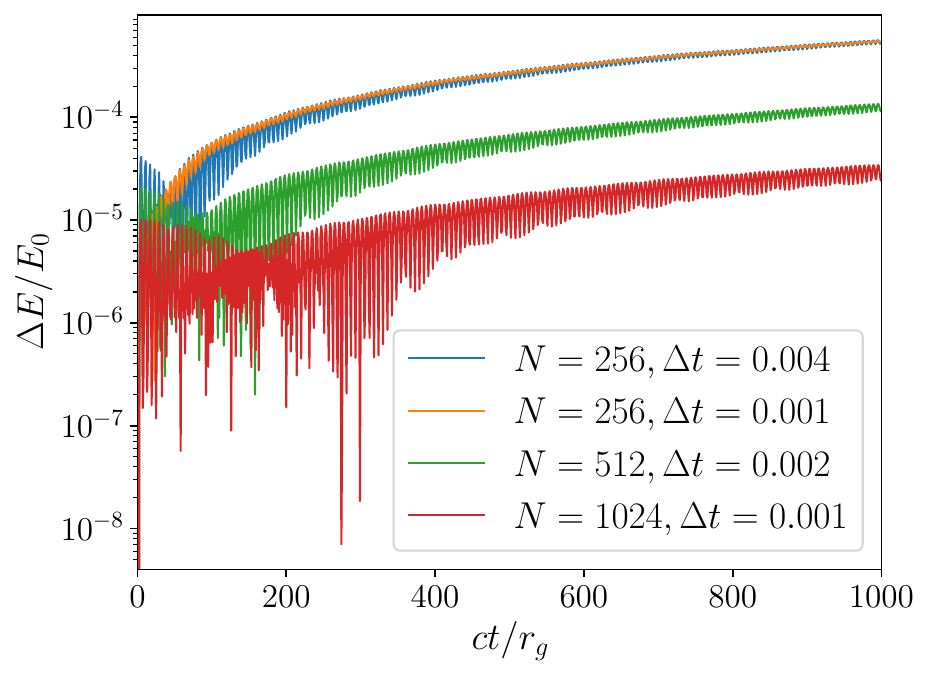}
    \caption{Deviation of energy from its original value for the charged particle orbit RKA3. $N = N_{r} = N_{\theta}$ is the number of grid points in each dimension, and $\Delta E = |E - E_{0}|$ is the deviation of the particle energy from its initial value. The energy error remains the same when only decreasing $\Delta t$, indicating that it is dominated by field interpolation error.}
    \label{fig:EvsDxDt}
\end{figure}

\begin{figure*}[t!]
    \centering
    \includegraphics[width=\textwidth]{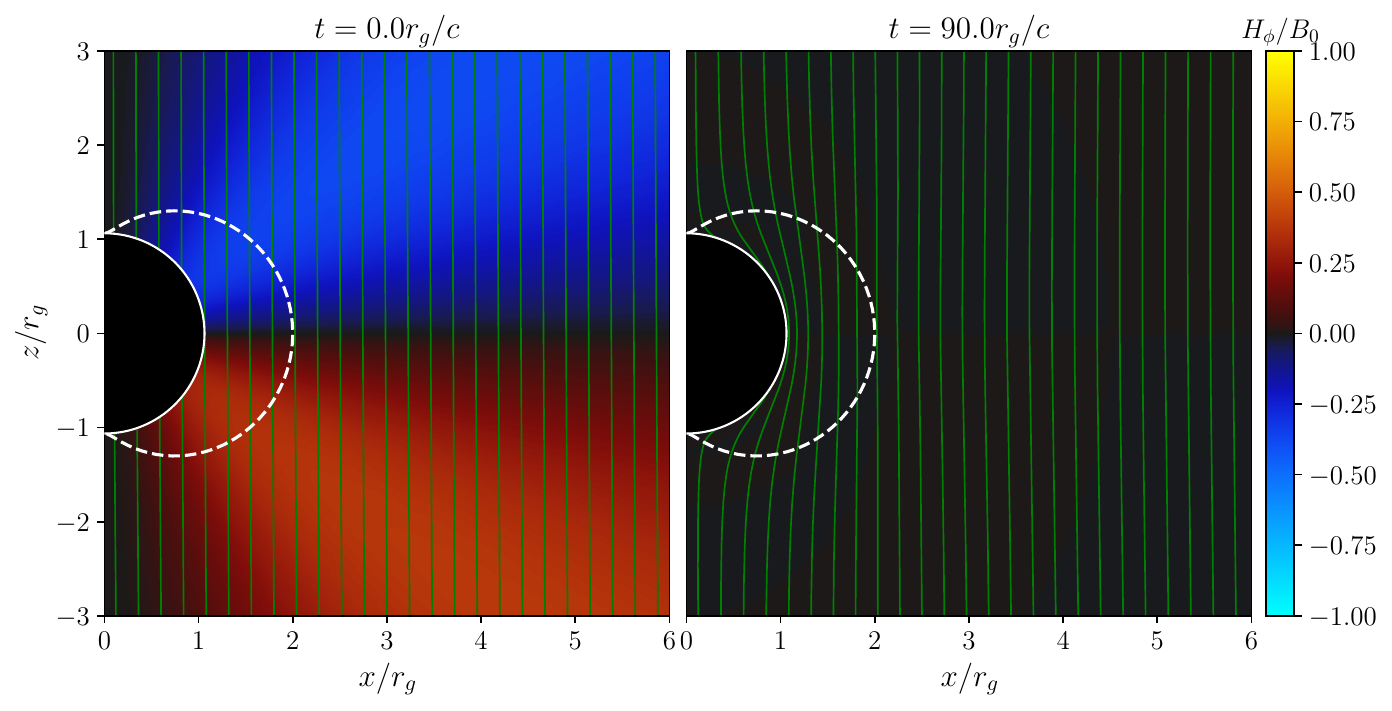}
    \caption{Vacuum Wald solution for black hole spin $a = 0.998$. \emph{Left:} Initial state, which is the Wald solution for a non-rotating black hole; \emph{Right:} Steady state after $90r_{g}/c$. The green lines depict magnetic field lines, and color plot shows the value of $H_{\phi}/B_{0}$. The solid black sphere shows the horizon of the black hole and the dashed white curve shows the ergosphere. In the steady state, most of the magnetic flux is expelled from the horizon, and $H_{\phi}$ goes to zero. }
    \label{fig:vacuum-wald}
\end{figure*}

We also examined the energy conservation properties of the \aperture{} particle integrator in these charged particle trajectories. In the presence of the electromagnetic field, the conserved energy is modified by the 0-th component of the vector potential:
\begin{equation}
    \label{eq:ptc_energy}
    \begin{split}
    E=-\pi_0=-u_0-\frac{q}{m} A_0.
    \end{split}
\end{equation}
Even though the geodesic integrator by itself is symplectic, and the Boris pusher generally conserves energy~\citep[see e.g.][]{2013PhPl...20h4503Q}, the complete algorithm incurs a finite energy error at each step. Figure~\ref{fig:EvsDxDt} shows the deviation of energy from its original value for the RKA3 orbit. We generally see a linear increase of $\Delta E$ with respect to time, with overall error decreasing with $\Delta r$ and $\Delta t$. We believe this error can be attributed to the interpolation of field values from the grid points to the particle location, especially when using first order interpolation. Due to this error that is related to $\Delta r$ and $\Delta \theta$, we observe identical energy error growth when we decrease $\Delta t$ but keep the number of grid cells fixed, as shown in Figure~\ref{fig:EvsDxDt}. To reduce the energy error, one needs to reduce the grid spacing $\Delta r$ and $\Delta \theta$, as well as the time step size $\Delta t$ at the same time.


Since a large portion of the error in the full particle pusher arises from
  field interpolation, utilizing a higher order particle integrator is of
  limited value. For typical simulation resolutions
  ($N_{r}\approx N_{\theta} \gtrsim 1000$) and time step sizes
  $\Delta t \lesssim 10^{-3}$, the long-term cumulative relative error is on the order of $10^{-5}$. We believe this level of error in the particle update
  algorithm is adequate for the purpose of large-scale PIC
  simulations.

\subsection{Vacuum Wald Solution}
\label{sec:vacuum-wald}

To test the correctness of our field solver, we simulate a spinning black hole in the presence of a uniform magnetic field in vacuum. The steady state of this configuration is exactly solvable and was found by~\citet{1974PhRvD..10.1680W}. In our test case, we start with a Wald solution for a non-rotating black hole ($a = 0$) in a Kerr spacetime with $a = 0.998$. We use the equations given by~\citet{2004MNRAS.350..427K}:
\begin{equation}
    \label{eq:vacuum-nonrotating-wald}
    E_{i} = 0,\quad B^{i} = \frac{B_{0}}{2\sqrt{\gamma}}(0, -\partial_{\theta}\gamma_{\phi\phi}, \partial_{r}\gamma_{\phi\phi}),
\end{equation}
which implies that $D^{\phi}$ is nonzero and needs to be solved using Equation~\eqref{eq:E-H-B-D}. 


\begin{figure*}[t]
    \centering
    \includegraphics[width=\textwidth]{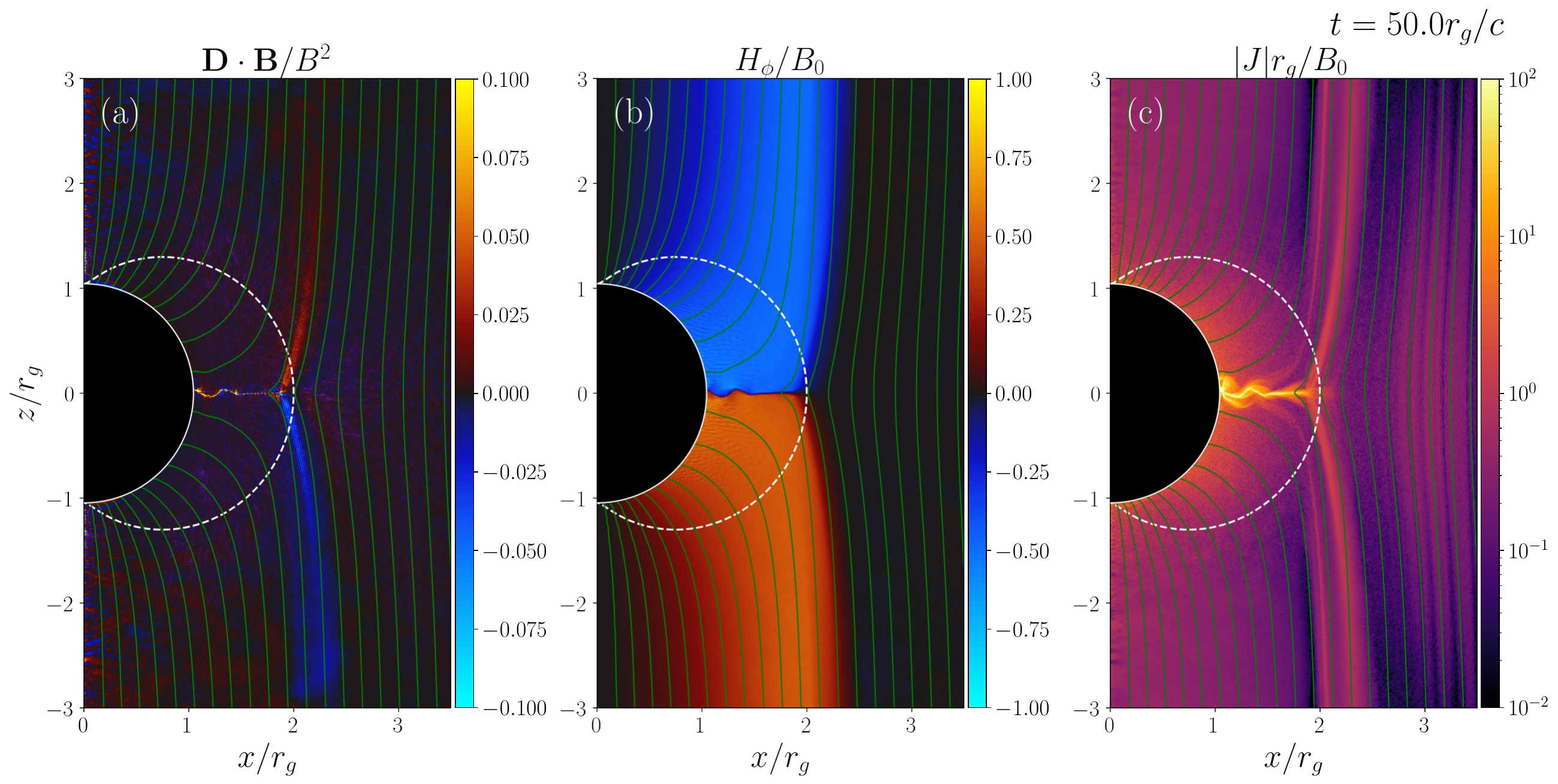}
    \caption{Plasma-filled Wald solution for black hole spin $a = 0.999$. From left to right are: (a) parallel electric field normalized to the local magnetic field; (b) value of $H_{\phi}$ normalized to the strength of the uniform magnetic field $B_{0}$; (c) magnitude of the 3-current $J$. The green lines depict magnetic field lines, the black sphere with solid white outline denotes the black hole horizon, and the dashed white curve marks the ergosphere. The current sheet within the ergosphere is clearly seen and becomes kink unstable. }
    \label{fig:plasma-wald}
\end{figure*}

After starting the simulation, due to the mismatch of the solution with the
actual black hole spin, the electromagnetic field is expected to evolve and
establish a new steady state which agrees with the spinning Wald solution. The
system achieves the transition through a series of damped oscillations near the
event horizon. Figure~\ref{fig:vacuum-wald} shows the initial state and final
steady state of our vacuum Wald test case. The initial state of the non-rotating
solution given by Equation~\eqref{eq:vacuum-nonrotating-wald}, when inserted
into a spinning Kerr metric, leads to nonzero $H_{\phi}$ everywhere in the
simulation domain. After many oscillations, the magnetic field configuration is
indeed able to self-consistently evolve to a steady state that has
$H_{\phi} = 0$ and agrees with the spinning Wald solution. In this state, most
of the magnetic flux that originally threads the horizon becomes expelled, which
is often called the ``Meissner effect'' for rotating black holes~\citep[see
e.g.][]{PhysRevD.12.3037}. This test case demonstrates that our field solver
described in Section~\ref{sec:field-solver} is well-behaved over a long time, in
the absence of source terms.


\subsection{Plasma-filled Wald Solution}
\label{sec:plasma-wald}

As a final test, we attempt to reproduce the results
by~\citet{2019PhRvL.122c5101P}, simulating a plasma-filled magnetosphere of a
spinning black hole that is threaded by a uniform magnetic field using \aperture{}. We start from the vacuum rotating Wald solution (detailed in Appendix~\ref{app:monopole}), which is depicted in the right panel of Figure~\ref{fig:vacuum-wald}. We inject $e^{\pm}$ pairs whenever local magnetization is larger than a threshold value, $\sigma_\mathrm{thr}$ and when local $\bm{D}\cdot\bm{B}/B^{2}$ exceeds a critical value $\epsilon_{D\cdot B}$.

Figure~\ref{fig:plasma-wald} shows a simulation of the plasma-filled Wald
problem. The simulation uses a box of resolution
$N_{r}\times N_{\theta} = 1024\times 1024$ with logarithmic spacing in $r$. The
magnetic field strength at infinity is $B_{0} = 500 m_{e}c^{2}/e r_{g}$ and
black hole spin is $a = 0.999$. We use a critical injection threshold of
$\epsilon_{D\cdot B} = 10^{-3}$, similar to the ``high plasma supply'' scenario
described by~\citet{2019PhRvL.122c5101P}. We observe that, as a result of the injection of plasma, magnetic field lines that thread the ergosphere are dragged into the black hole. This forms an equatorial current sheet inside the ergosphere, much like the equatorial current sheet of a rotating pulsar. This current sheet undergoes magnetic reconnection and also becomes kink unstable.
Note that, due to the nature of our 2D axisymmetric simulation, the toroidal magnetic field cannot reconnect through the tearing instability, as the $\phi$ direction is not resolved. Near the black hole horizon, the magnetic field is far dominated by $B_{\phi}$, therefore a significant part of the potential kinetic physics related to the current sheet in the ergosphere can only be included in a full 3D simulation.


\begin{figure}[h]
    \centering
    \includegraphics[width=0.48\textwidth]{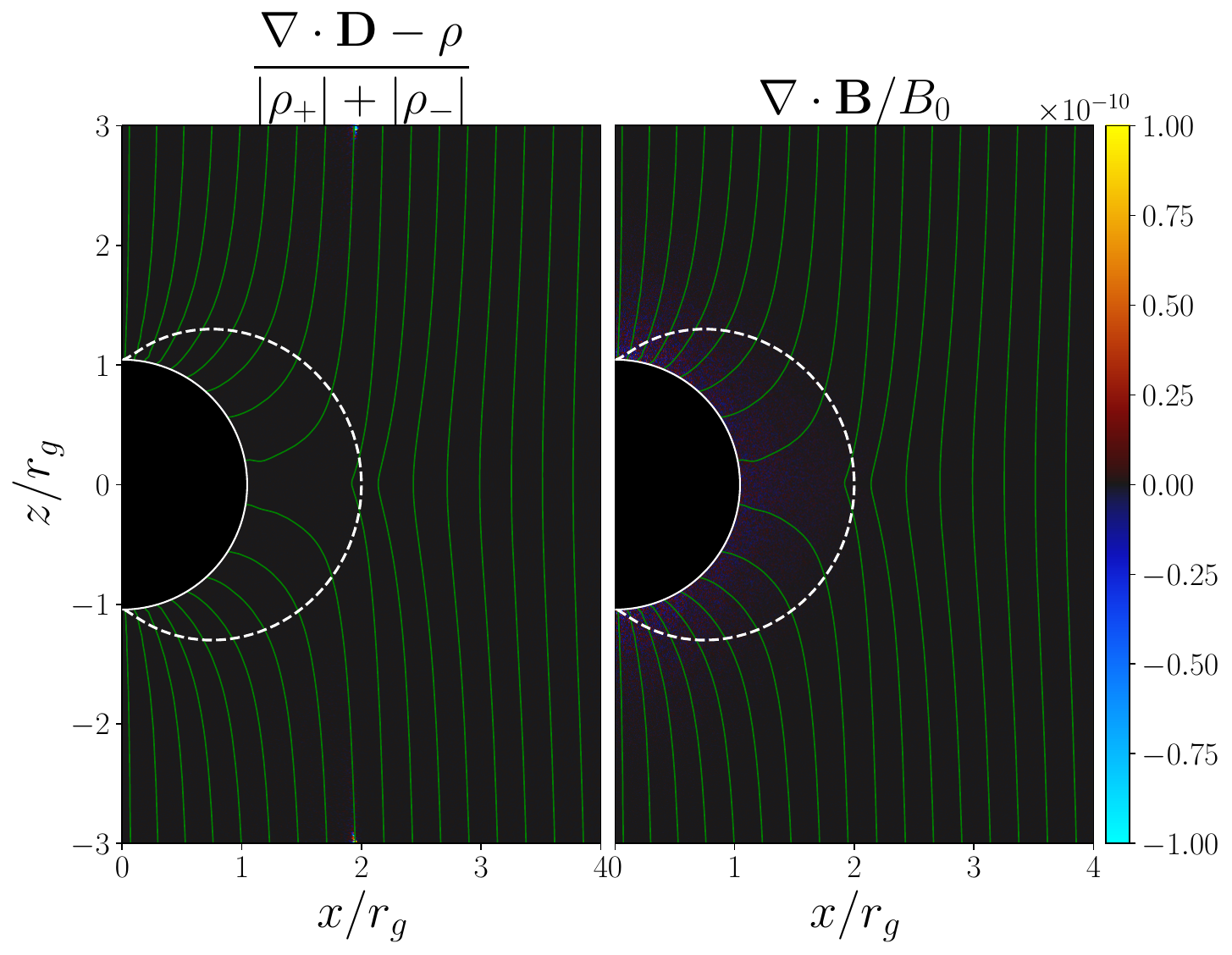}
    \caption{Divergence errors at the end of the plasma-filled Wald simulation, $t=50r_{g}/c$. The cumulative relative error in both $\nabla\cdot \bm{D}$ and $\nabla\cdot \bm{B}$ are everywhere less than $10^{-10}$ without the need for divergence cleaning.}
    \label{fig:divD-error}
\end{figure}

It is instructive to check the extent of charge conservation of the algorithm described in Section~\ref{sec:current-deposition}. Figure~\ref{fig:divD-error} shows the error of both $\nabla\cdot \bm{D}$ and $\nabla\cdot\bm{B}$ at the end of the simulation, after 50,000 time steps. The relative cumulative error on both quantities are on the $10^{-11}$ level, implying that the truncation error per time step is on the order of \verb+double+ precision floating point rounding error. This demonstrates numerical charge conservation up to machine precision without the need of divergence cleaning.


\section{Spark Gaps in a Monopolar Magnetic Field}
\label{sec:application}

As a concrete application of \aperture{}, we perform a series of numerical
experiments on spark gaps in the black hole magnetosphere. This problem has been
studied before using GRPIC simulations in 1D~\citep[e.g.][]{2018A&A...616A.184L,
  2020ApJ...895..121C, 2020ApJ...902...80K} and 2D~\citep{2020PhRvL.124n5101C,
  2023MNRAS.526.2709N}. Curiously, the results in 1D and 2D presented in the
literature are qualitatively different. One-dimensional simulations tend to find
a solution where a well-defined, macroscopic (size comparable to~$r_{g}$) ``gap''
region of $\bm{D}\cdot \bm{B}\neq 0$ opens and closes quasi-periodically
near the null charge surface, accelerating charges and igniting pair production
on a limit cycle. Two-dimensional simulations, however, tend to find a solution
where the acceleration and pair production regions are small and stochastically
distributed in a broad region close to the black hole. No large-scale coherent
``gap''-like structure has been reported in 2D GRPIC simulations so far, and no
clear physical explanation for this discrepancy with 1D results has yet been put
forward. One potential idea was that the gap dynamics is determined by the frame where radiative transfer computation is performed. For example, the 1D simulations typically assumed an isotropic soft photon field in the ZAMO frame, whereas the 2D simulations typically assumed the same isotropic soft photon field but in the Kerr-Schild FIDO frame.

\begin{figure*}[t]
    \centering
    \includegraphics[width=\linewidth]{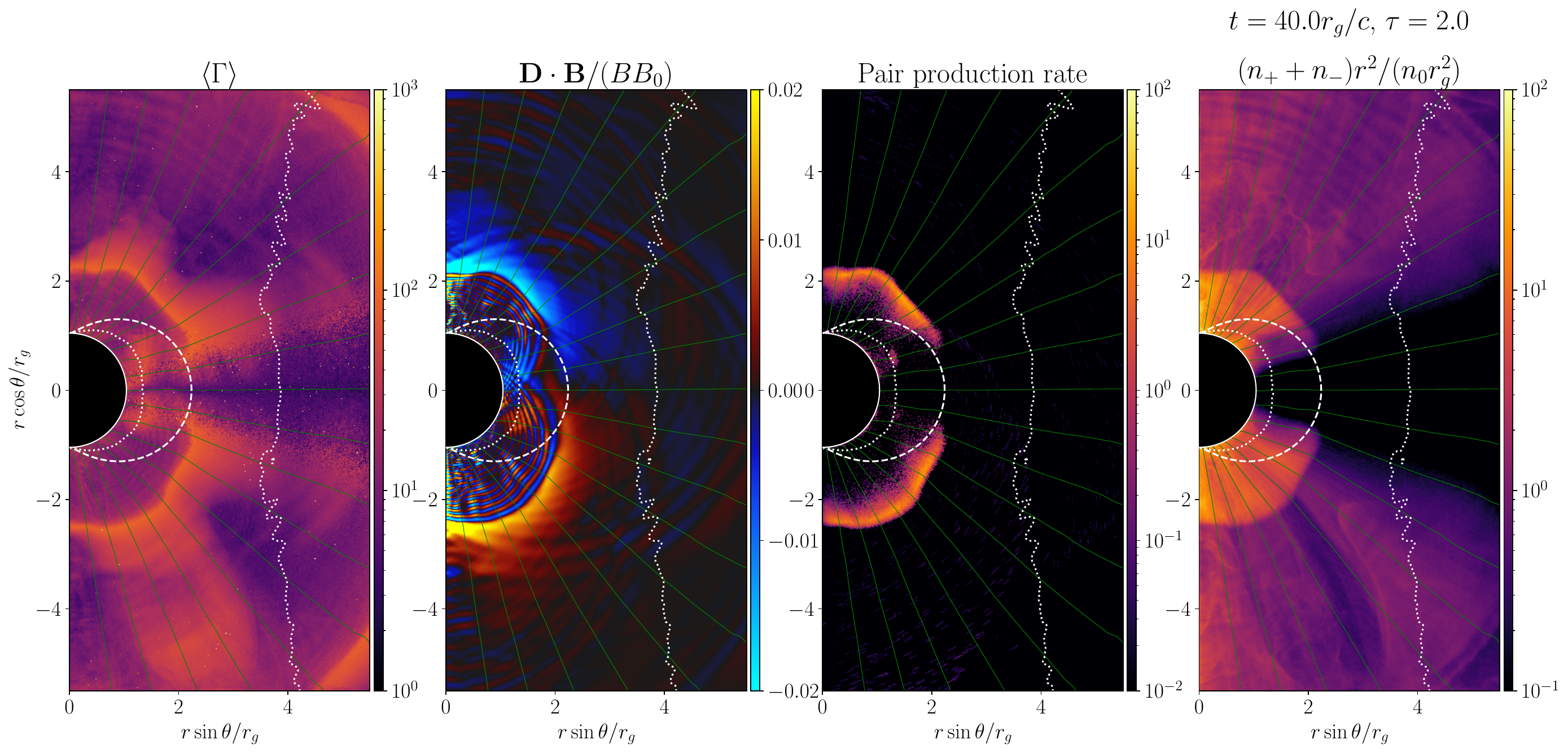}
    \caption{Snapshots from a simulation of gaps in a monopolar magnetosphere. The simulation has $B_0=2\times 10^3 m_ec^2/(er_g)$ and opacity $\tau=2.0$ (our fiducial run). From left to right: average Lorentz factor of particles in the FIDO frame $\langle \Gamma\rangle$, the electric field component parallel to the magnetic field $\bm{D}\cdot\bm{B}/(B B_0)$, pair production rate (we plot the value of $r\sqrt{-g}\, (dS^0/dt)$ in arbitrary units, where $S^0$ is the zeroth component of the number-flux-4-vector), and the plasma density. Green lines are magnetic field lines. The white dashed line is the ergosphere, and the white dotted lines are the light surfaces.}
    \label{fig:monopole_2D}
\end{figure*}

As an attempt to resolve this discrepancy, we apply the \aperture{} code to study this problem using a simplified prescription
for pair production, with the hope to disentangle how the large-scale
electrodynamics near the black hole is affected self-consistently by the details
of pair production microphysics. We use a prescription inspired by pulsar
simulations such as by~\citet{2014ApJ...795L..22C}, where a photon would be
emitted once an electron or positron reaches a threshold Lorentz factor $\gamma_\mathrm{thr}$
in the FIDO frame. The emitted photon has a fixed amount of energy in the local
FIDO frame $\varepsilon_\mathrm{ph}$, and moves in the same direction as the
emitting particle. This photon is then propagated in the code using the photon
equations of motion~(Equations~\eqref{eq:particle-eom} but without the Lorentz
force $F_i$, and $u^0$ is computed using $\epsilon = 0$). We define an opacity
$\tau$ for the photon to convert into an $e^\pm$ pair. At every time step, the
probability for the photon to convert is given by the optical depth:
\begin{equation}
    \label{eq:optical-depth}
    P_{\gamma\to\pm} = 1 - e^{-\tau \Delta t} \approx \tau \Delta t.
\end{equation}
The approximation is valid when $\tau \Delta t \ll 1$, which is always the case for our simulations.

\begin{figure}
    \centering
    \includegraphics[width=\columnwidth]{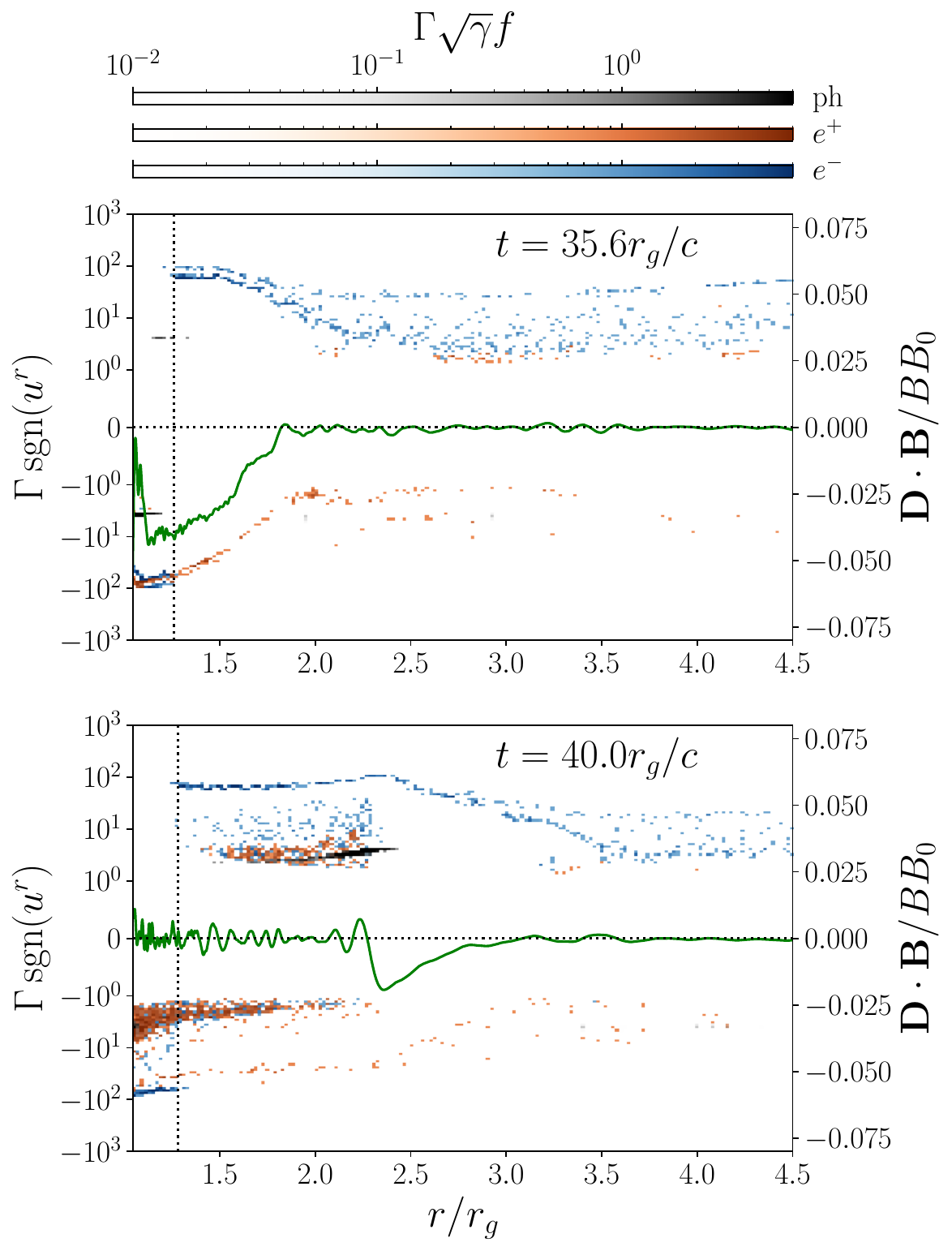}
    \caption{The phase space distribution of electrons (blue), positrons (orange) and photons (gray) along a radial line at $\theta=35^{\circ}$, at two different times during a cycle of gap opening and screening in our fiducial run (the same run as in Figure \ref{fig:monopole_2D}). The energy distribution is in terms of the particle Lorentz factor $\Gamma$ in the FIDO frame (for photons, we use $\epsilon/(m_ec^2)$, where $\epsilon$ is the photon energy in the FIDO frame), and we separate outgoing particles and ingoing particles using the sign of the radial component of their 4-velocity $u^r$. Also plotted is the normalized parallel electric field $\bm{D}\cdot\bm{B}/B B_0$ (the green line). The vertical dotted line marks the location of the inner light surface. The left boundary of the horizontal axis corresponds to the event horizon.}
    \label{fig:monopole_1D_phase}
\end{figure}

Our simulations start with a vacuum magnetic monopole solution (see Appendix~\ref{app:monopole}), similar to the simulations by~\citet{2020PhRvL.124n5101C}. We initialize the domain with a pair plasma, which would then start to move under the influence of the electromagnetic field. We performed simulations for a range of different pair production opacities $\tau\in [0.2, 2.0, 20]$, with the same magnetic field strength $B_0=2\times 10^3 m_ec^2/(er_g)$. The simulation resolution is fixed at $N_r\times N_\theta = 2048\times 2048$ with logarithmic spacing in $r$, and our time step size is chosen conservatively at $\Delta t = 5\times 10^{-4}r_{g}/c$. This resolution sufficiently resolves the Goldreich-Julian charge density near the event horizon, which can reach ${\sim}10\,B_{0}/4\pi r_{g}$ near the horizon according to numerical Force-free solutions~\citep{2020ApJ...895..121C}. We use the same threshold Lorentz factor $\gamma_\mathrm{thr} = 100$ for all of these simulations. Once a particle reaches the energy threshold, it emits a photon of energy $\varepsilon_\mathrm{ph} = 4.0$ measured in the FIDO frame. When converting to an $e^\pm$ pair, each particle gets exactly half of the original photon energy, which may have changed as the photon propagates in the spacetime.


\begin{figure*}
    \centering
    \includegraphics[width=0.98\textwidth]{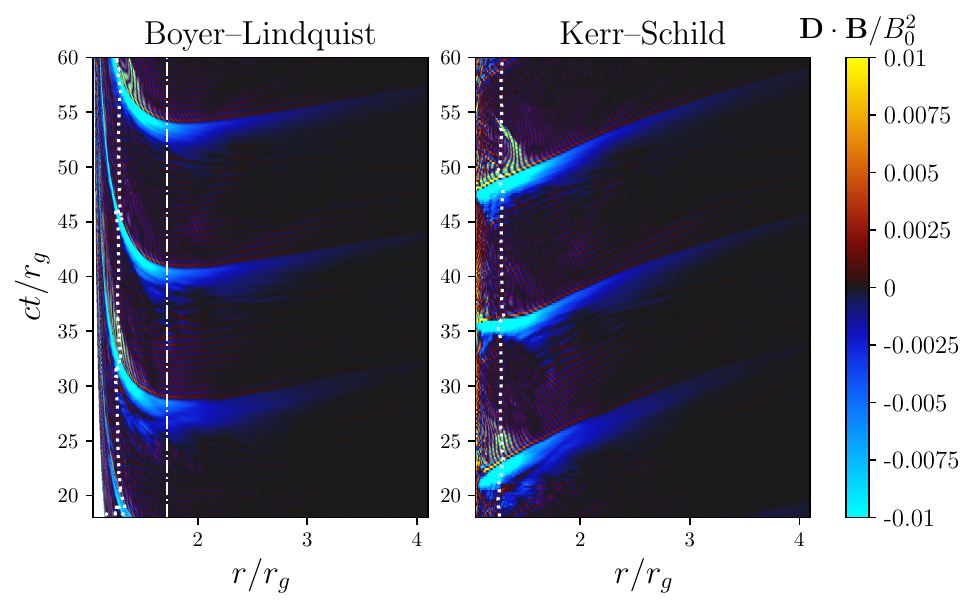}
    \caption{Spacetime diagram of $\mathbf{D}\cdot\mathbf{B}/B_0^2$ along a radial line at $\theta=35^{\circ}$ for our fiducial run, in Boyer-Lindquist coordinates (left) and Kerr-Schild coordinates (right), respectively. The white dotted line in each panel corresponds to the inner light surface. The white dash-dotted line in the left panel marks the location of the null surface in Boyer-Lindquist coordinates.}
    \label{fig:spacetime_1D}
\end{figure*}

In all simulations, we see quasiperiodic opening and screening of macroscopic spark gaps in the black hole magnetosphere. We first present the results from a fiducial run with opacity $\tau=2.0$. A cycle starts when an initially screened magnetosphere loses plasma as the particles fall into the black hole or get flung out at large radii. This leads to the development of regions with nonzero electric field $\bm{D}$ parallel to the magnetic field $\bm{B}$ close to the event horizon in both hemispheres. These gaps expand in size and the electric field grows with time. Particles are accelerated to reach the threshold energy, initiating pair production. This process first starts somewhere between the event horizon and the inner light surface, and the screening of the gap also starts around there. The screening process shows the typical large amplitude electrostatic oscillations. During the screening, the front of the gap keeps expanding out, and the screening front follows behind. As a result, it appears as if the gaps are moving outward.

Figure~\ref{fig:monopole_2D} shows 2D snapshots from the simulation when the gaps are active. We can see that $\bm{D}\cdot\bm{B}$ in the gap is negative in the northern hemisphere and positive in the southern hemisphere. Since the monopole magnetic field points outward, the direction of the gap electric field is consistent with the direction of the current flowing along the magnetic field (the current flows outward in the southern hemisphere and inward in the northern hemisphere). We also see that there is not much pair production near the equator. This is because the current density goes to zero at the equator. This result confirms that the formation of the gap is ultimately a current-driven kinetic phenomenon~\citep[see e.g.][]{2008ApJ...683L..41B}, instead of driven by the diverging MHD flow~\citep[see e.g.][]{2015ApJ...809...97B}. Near the equator, the centrifugal effect is actually the strongest, however it fails to drive pair discharge along these field lines.

In Figure~\ref{fig:monopole_1D_phase}, we show two snapshots of the phase space distribution of particles along a radial line at $\theta=35^{\circ}$, during one of the gap screening cycles. In the top panel, the gap has just opened up and the parallel electric field has reached the maximum. We can see that the gap electric field accelerates electrons and positrons in opposite directions. Note that although the gap can straddle the inner light surface, particles confined to magnetic field lines can only move inward inside the inner light surface, hence the change of ${\rm sgn}(u^r)$ at the inner light surface. In the bottom panel, the gap has mostly been screened, and the front of the gap has expanded outward. We can clearly see the population of pairs newly produced; these secondary electrons and positrons move in both directions. The macroscopic gap has size comparable to $r_{g}$, and emerges almost instantaneously. As a result, it is not clear whether it is developed from the inner light surface or from the event horizon itself.

To properly compare with earlier 1D GRPIC simulations, e.g., \citet{2020ApJ...895..121C}, our results need to be presented in the same coordinate system. \citet{2020ApJ...895..121C} used Boyer-Lindquist coordinates, while the 2D GRPIC simulations are usually done in Kerr-Schild coordinates (including this work). Although the $r$ and $\theta$ coordinates are the same between the two systems, $t$ and $\phi$ are different: they transform following Equation (\ref{eq:BL_KS_diff}) and (\ref{eq:BL_KS_global}) in Appendix \ref{app:kerr-schild}. In particular, the time coordinate involves a nonlinear transformation mixing $t$ with radius $r$. Therefore, time dependent features can have different appearance when using the different coordinate systems. Since the spark gap is inherently a time-dependent phenomenon, it is conceivable that its evolution may look different depending on the coordinate system used.

We have carried out the global transformation between the two coordinate systems on the results of our fiducial run. Figure~\ref{fig:spacetime_1D} shows a 1D spacetime diagram of $\bm{D}\cdot\bm{B}/B_0^2$, along a radial line at $\theta=35^{\circ}$, in both Boyer-Lindquist coordinates and Kerr-Schild coordinates. The quantity $\bm{D}\cdot\bm{B}$ is a Lorentz scalar, so its value does not change upon the coordinate transformation, but the nonzero regions may get remapped into different spacetime locations. In Kerr-Schild coordinates, the gaps start to develop near the event horizon and the inner light surface, then move outward to larger radii, at a speed of $\sim 0.25c$. In Boyer-Lindquist coordinates, the time slicing is different, and we see the gaps start somewhere in between the two light surfaces---the location turns out to be consistent with the null surface, where the Boyer-Lindquist FIDO (also the zero angular momentum observer, or ZAMO) measured charge density is zero. The gaps then expand both inward and outward; the screening also starts near the null surface and expands in both directions.

\begin{figure*}[t]
    \centering
    \includegraphics[width=0.9\textwidth]{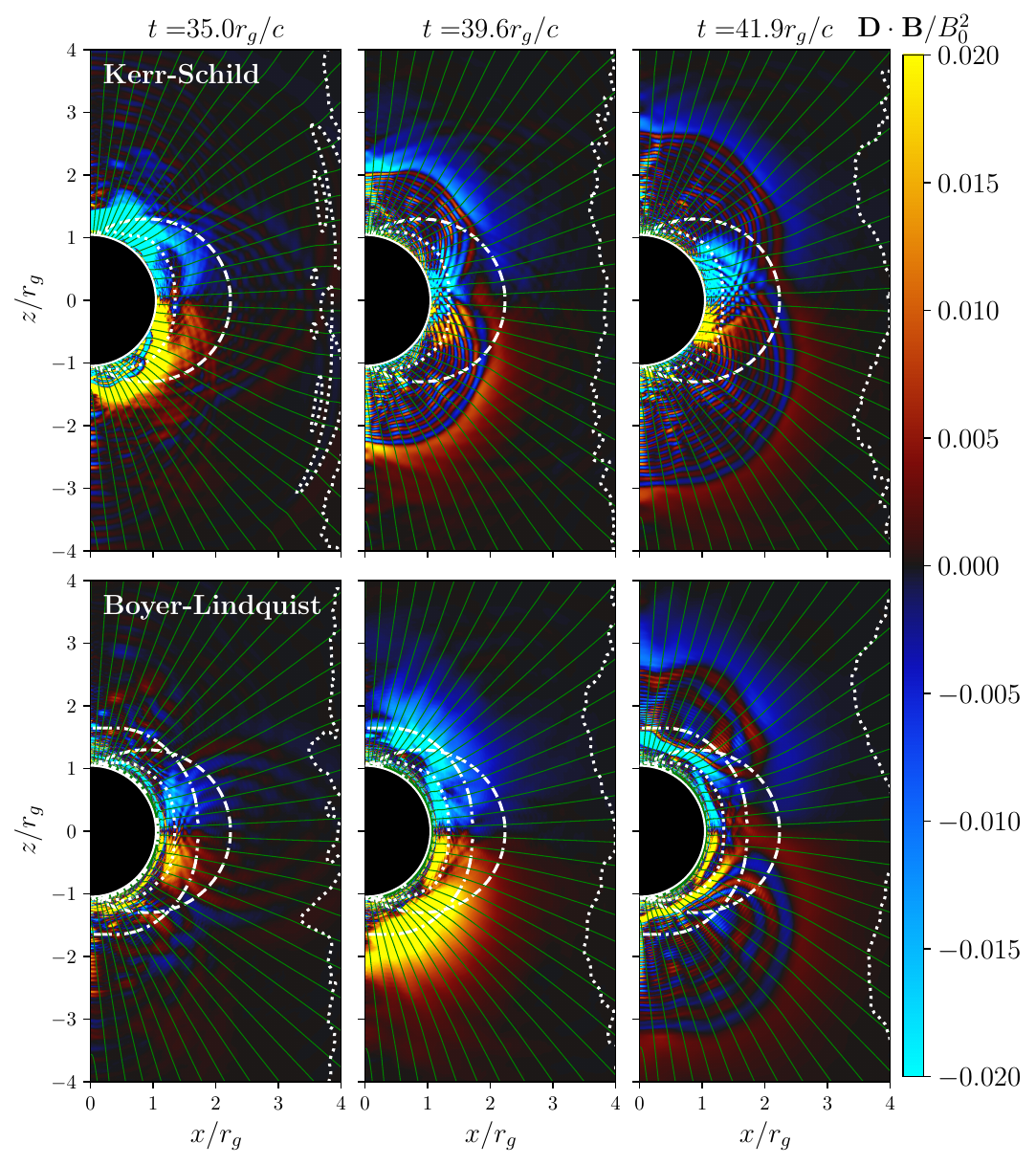}
    \caption{2D snapshots of $\mathbf{D}\cdot\mathbf{B}/B_0^2$ at constant coordinate times for our fiducial run, in Kerr-Schild coordinates (top row) and Boyer-Lindquist coordinates (bottom row). Green lines are poloidal magnetic field lines. The white dashed line is the ergosphere, the white dotted lines are the light surfaces, and the white dash-dotted line in the bottom row corresponds to the null surface in Boyer-Lindquist coordinates, which is not present in Kerr-Schild coordinates.}
    \label{fig:BL-vs-KS}
\end{figure*}

\begin{figure*}[t]
    \centering
    \includegraphics[width=0.98\textwidth]{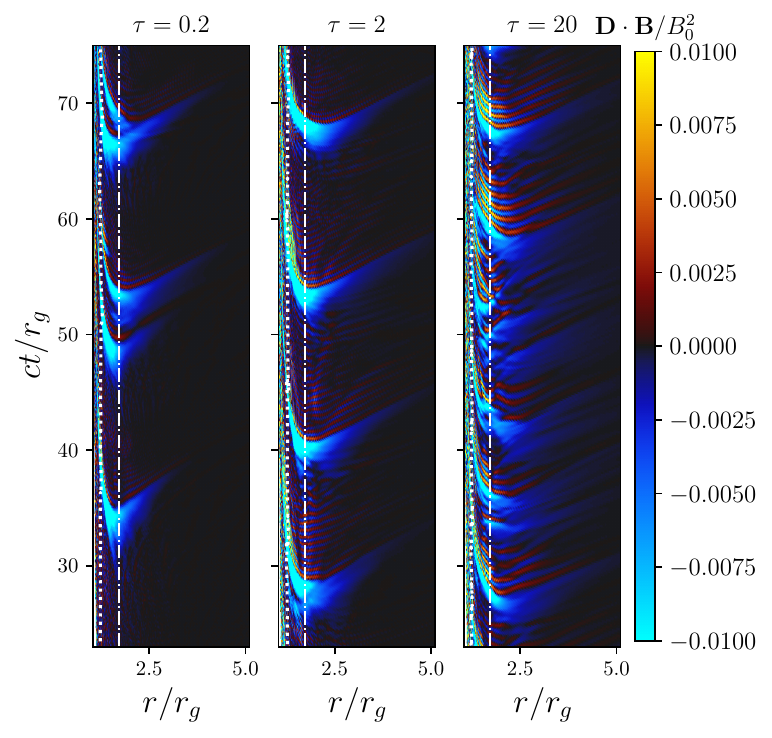}
    \caption{A comparison of the spacetime evolution of the gap in Boyer-Lindquist coordinates along $\theta=35^{\circ}$ for a range of different pair production opacity $\tau=0.2, 2, 20$. Color shows the Lorentz scalar $\mathbf{D}\cdot\mathbf{B}/B_0^2$. The white dotted line is the inner light surface and the white dash-dotted line is the null surface.}
    \label{fig:opacity_comparison}
\end{figure*}

We also show a few 2D snapshots at constant coordinate times for Kerr-Schild and Boyer-Lindquist coordinate systems, respectively, in Figure~\ref{fig:BL-vs-KS}. Again, in Kerr-Schild coordinates, the gaps start
somewhere between the event horizon and the inner light surface, then move outward while being screened. In Boyer-Lindquist coordinates, the gaps develop and first get screened near the null surface, at a latitude where the electric field peaks; the large scale oscillation associated with the screening appears as concentric oval rings that expands out in all directions. It takes an infinitely long amount of coordinate time for features to move toward the event horizon and fall into the black hole, consistent with the properties of the Boyer-Lindquist coordinates.

These results show that the gap dynamics we observe in the 2D GRPIC simulations are consistent with earlier 1D works of \citet{2020ApJ...895..121C} and \citet{2020ApJ...902...80K}. Although the gap opening and screening are physical events, their appearance can be different when
viewed by different observers, which in our case correspond to different coordinate systems
with different time slicing. The null surface is a frame dependent feature by definition; it shows up for ZAMO, or Boyer-Lindquist FIDO observers, but not the Kerr-Schild FIDO: the charge density measured by Kerr-Schild FIDO does not change sign along magnetic field lines. Our results suggest that it is indeed physical for Boyer-Lindquist FIDO to observe gaps developing around the null surface, and this is also consistent with electrodynamics requirements, as we further discuss in \S\ref{sec:discussions}.

Next, we compare the results from a series of simulations with different pair production opacity $\tau$, while the magnetic field $B_0$ and the threshold Lorentz factor for photon emission $\gamma_{\rm thr}$ are held the same. We find that as $\tau$ increases, the gap size in the direction along the magnetic field becomes smaller, so does the pair production region. Gaps become less coherent in the transverse direction, and the time interval between successive gaps gets shorter. In Figure \ref{fig:opacity_comparison}, we show the 1D spacetime diagram of $\bm{D}\cdot\bm{B}/B_0^2$ along a radial line at $\theta=35^{\circ}$ in Boyer-Lindquist coordinates, for three runs with different opacity $\tau$. We can clearly see the different duty cycles. Nevertheless, we find that the overall dissipation level is similar for the three runs. This may suggest that the dissipation is mainly controlled by the threshold Lorentz factor $\gamma_{\rm thr}$ in this simplified setup. This comparison is mainly of academic interest, since our pair production prescription is highly artificial. The proper dependence of the microphysics on the optical depth to inverse Compton scattering will be explored in a parallel work by Yuan et al. (in preparation).

\section{Discussions}
\label{sec:discussions}

This paper presented a general relativistic Particle-in-Cell (GRPIC) code framework \aperture{}, and all of the relevant numerical algorithms it employs. We included a list of test cases to validate the correctness and convergence of various aspects of the code, including test particle trajectories, vacuum field solution, and a plasma-filled Wald solution. It was verified that the code observes local current conservation down to numerical precision. We then applied the code to study a simple model of pair discharge in the black hole magnetosphere, and find that the electrodynamics resemble the previously reported 1D quasi-periodic solutions~\citep[e.g.][]{2020ApJ...895..121C}, especially after performing a nonlinear global transformation to the Boyer-Lindquist coordinates. A large-scale ``gap'' region with coherent $\bm{D}\cdot\bm{B}$ can develop around the null charge surface in Boyer-Lindquist coordinates, and become screened by $e^\pm$ pairs produced \emph{in situ}. On the other hand, due to the different time slicing in Kerr-Schild coordinates, the dynamics appears different there. The gap develops and becomes screened from close to the event horizon and the inner light surface, leading to a gap that moves outwards as it is being screened. This is in general agreement with the previously reported gap dynamics in 2D~\citep{2020PhRvL.124n5101C}. Our results should resolve the outstanding discrepancy in the literature regarding the gap dynamics, and provide a unified framework to analyze this type of physical process.

The origin of the recurrent gaps is very similar to the mechanism of ``outer gaps'' in pulsar magnetospheres, proposed by~\citet{1986ApJ...300..500C}. The requirement for such a gap to develop is the existence of a null surface, across which the background charge density changes sign. In addition, the magnetospheric current needs to have the right sign to drive charge separation from the null surface when the pair multiplicity becomes low. The monopole black hole magnetosphere satisfies this requirement. Since the null surface is a coordinate-dependent construct, we conjecture that in any 3+1 slicing of the spacetime, if a null charge surface exists, then the gap should first develop there. The results presented in this paper supports this conjecture, but proving it for all 3+1 foliations will require further work that is beyond the scope of this numerical paper. In the absence of a null surface, such as in Kerr-Schild coordinates, a natural surface of separation is the inner light surface. This may be the reason that we see the gap developing and being screened close to the inner light surface, similar to the results reported by~\citet{2020PhRvL.124n5101C}.

In this paper, we have employed a simplified scheme for pair production, namely emitting photons at a fixed threshold energy and setting an energy-independent optical depth for pair production. A parallel work is being done (Yuan et al, in preparation) to study the fully self-consistent $e^\pm$ discharge problem in black hole magnetospheres using \aperture{}, taking into account physical cross sections for inverse Compton scattering and pair production. Preliminary results suggest that the qualitative features presented in this paper remain valid even when the full radiative processes are implemented. In particular, even when the radiative transfer is carried out in the FIDO of Kerr-Schild coordinates, the dynamics viewed in Boyer-Lindquist coordinates resemble what was reported in 1D simulations, where radiative transfer is carried out in the ZAMO. In other words, the simplified pair production scheme presented in this paper seems to already robustly capture the most important qualitative features of this problem, indicating that the particular angular distribution of the background photon field may not be as important as previously conjectured.

A GRPIC code opens the doors to many different studies of black hole magnetospheres in the collisionless regime that was inaccessible before. We believe in the near future, GPU powered GRPIC codes can definitively address some of the outstanding problems in low-luminosity AGN. An incomplete list of these problems include the energy partition between ions and electrons in the accretion flow, the source of heating and nonthermal particle acceleration in the accretion disk corona, and the time variability of the resulting radiation. We are excited to apply \aperture{} on some of these problems soon.

\begin{acknowledgments}

We thank Kyle Parfrey and Sam Gralla for helpful discussions. AC and YY acknowledge support from NSF grants DMS-2235457 and
AST-2308111. AC acknowledges additional support from NASA grant
80NSSC24K1095. YY also acknowledges support from the Multimessenger Plasma
Physics Center (MPPC), NSF grant PHY-2206608, and support from the Simons
Foundation (MP-SCMPS-00001470). This research used resources of the Oak
Ridge Leadership Computing Facility at the Oak Ridge National Laboratory,
which is supported by the Office of Science of the U.S. Department of Energy
under Contract No. DE-AC05-00OR22725.

\end{acknowledgments}

\appendix

\section{Kerr-Schild Coordinates}
\label{app:kerr-schild}

This paper follows~\citet{2004MNRAS.350..427K} in terms of notations for the Kerr metric and 3+1 formulation. The Kerr-Schild coordinate metric is given by:
\begin{equation}
    \label{eq:kerr-schild-metric}
    ds^{2} = g_{tt}\,dt^{2} + 2g_{t\phi}\,dt d\phi + 2g_{tr}\,dt dr + g_{rr}\,dr^{2} + g_{\theta\theta}\,d\theta^{2} + g_{\phi\phi}\,d\phi^{2} + 2g_{r\phi}\,dr d\phi.
\end{equation}
Notably, there are cross terms for $t\phi$, $tr$, and $r\phi$. The metric coefficients are (taking $M = 1$):
\begin{equation}
    \label{eq:ks-coefficients}
    \begin{split}
      g_{tt} &= z - 1,\quad g_{t\phi} = -az \sin^{2}\theta,\quad g_{tr} = z,\quad g_{rr} = 1 + z, \\
      g_{\theta\theta} &= \rho^{2},\quad g_{\phi\phi} = \Sigma \sin^{2}\theta / \rho^{2},\quad g_{r\phi} = -a \sin^{2}\theta (1 + z),
    \end{split}
\end{equation}
where:
\begin{equation}
    \label{eq:ks-variables}
    \begin{split}
      \rho^{2} &= r^{2} + a^{2}\cos^{2}\theta, \quad z = 2r / \rho^{2},\\
      \Sigma &= (r^{2} + a^{2})^{2} - a^{2}\Delta \sin^{2}\theta,\quad \Delta = r^{2} + a^{2} - 2r. \\
    \end{split}
\end{equation}

The spatial metric is simply given by $\gamma_{ij} = g_{ij}$, and its inverse is:
\begin{equation}
    \label{eq:gamma-inverse}
    \gamma^{ij} = \begin{pmatrix}
      \displaystyle\frac{a^{2} + r^{2}}{\rho^{2}} - \frac{2r}{\rho^{2} + 2r} & 0 & \displaystyle\frac{a}{\rho^{2}} \\
      0 & \displaystyle \frac{1}{\rho^{2}} & 0 \\
      \displaystyle \frac{a}{\rho^{2}} & 0 & \displaystyle \frac{1}{\rho^{2}\sin^{2}\theta}.
    \end{pmatrix}
\end{equation}
The lapse function and the shift vector for the Kerr-Schild metric are:
\begin{equation}
    \label{eq:ks-alpha-beta}
    \alpha = \frac{1}{\sqrt{1 + z}},\quad \beta^{r} = \frac{z}{1 + z},\quad \beta^{\theta} = \beta^{\phi} = 0.
\end{equation}
The coordinate observers of Kerr-Schild are on a series of infalling trajectories towards the black hole.

The transformation between Kerr-Schild and Boyer-Lindquist coordinates is the following:
\begin{equation}\label{eq:BL_KS_diff}
    \begin{split}
    dt_{KS}&=dt_{BL}+\frac{2r}{\Delta}dr_{BL},\\
    dr_{KS}&=dr_{BL},\\
    d\theta_{KS}&=d\theta_{BL},\\
    d\phi_{KS}&=d\phi_{BL}+\frac{a}{\Delta}dr_{BL}.
    \end{split}
\end{equation}

Upon integration, we obtain the global coordinate transformation \citep[e.g.,][]{2010MNRAS.406.2047G}
\begin{equation}\label{eq:BL_KS_global}
    \begin{split}
    t_{\rm KS}&=t_{\rm BL}+\frac{1}{\sqrt{1-a^2}}\left[\left(1+\sqrt{1-a^2}\right)\ln \left|\frac{r_{\rm BL}}{1+\sqrt{1-a^2}}-1\right|-\left(1-\sqrt{1-a^2}\right)\ln \left|\frac{r_{\rm BL}}{1-\sqrt{1-a^2}}-1\right|\right],\\
    r_{\rm KS}&=r_{\rm BL},\\
    \theta_{\rm KS}&=\theta_{\rm BL},\\
    \phi_{\rm KS}&=\phi_{\rm BL}+\frac{a}{2\sqrt{1-a^2}}\ln\left|\frac{r_{\rm BL}-(1+\sqrt{1-a^2})}{r_{\rm BL}-(1-\sqrt{1-a^2})}\right|.
    \end{split}
\end{equation}

\section{Discretized Field Equations}
\label{app:field-eqn-discrete}

In this appendix we present the fully discretized version of Equation~\eqref{eq:integral-form} and how we solve it in \aperture{}. The update equations for $D^{i}$ are:
\begin{equation}
    \label{eq:D-update}
    \begin{split}
      \frac{\Delta D^{r}_{ij}}{\Delta t} &= \frac{\left(H_{\phi, i, j} - H_{\phi, i, j-1}\right)\Delta \phi}{A^{D}_{r,ij}} - J^{r}_{ij} = C_{H}^{r} - J_{ij}^{r}\\
  \frac{\Delta D^{\theta}_{ij}}{\Delta t} &= \frac{\left(H_{\phi, i-1, j} - H_{\phi, i, j}\right)\Delta\phi}{A^{D}_{\theta,ij}} - J^{\theta}_{ij} = C_{H}^{\theta} - J_{ij}^{\theta} \\
  \frac{\Delta D^{\phi}_{ij}}{\Delta t} &= \frac{\left(H_{\theta, i, j} - H_{\theta, i-1, j}\right)\Delta\theta + \left(H_{r, i, j-1} - H_{r,i,j}\right)\Delta r}{A^{D}_{\phi,ij}} - J^{\phi}_{ij} = C_{H}^{\phi} - J_{ij}^{\phi},
    \end{split}
\end{equation}
where $i$ and $j$ are the grid indices in the $r$ and $\theta$ directions, respectively. The auxiliary field components $H_{i}$ are computed using:
\begin{equation}
    \label{eq:auxH}
    \begin{split}
      H_{r,ij} &= \alpha \gamma_{rr}B^{r}_{ij} + \frac{1}{2}\left((\alpha \gamma_{r\phi}B^{\phi})_{ij} + (\alpha \gamma_{r\phi}B^{\phi})_{i-1,j}\right)\\
  H_{\theta,ij} &= \alpha \gamma_{\theta\theta}B^{\theta}_{ij} + \frac{1}{2}\left((\sqrt{\gamma}\beta^{r}D^{\phi})_{ij} + (\sqrt{\gamma}\beta^{r}D^{\phi})_{i+1,j}\right) \\
  H_{\phi,ij} &= \alpha \gamma_{\phi\phi}B^{\phi}_{ij} + \frac{1}{2}\left((\alpha \gamma_{r\phi}B^{r})_{ij} + (\alpha \gamma_{r\phi}B^{r})_{i+1,j}\right) \\
      &\quad - \frac{1}{2}\left((\sqrt{\gamma}\beta^{r}D^{\theta})_{ij} + (\sqrt{\gamma}\beta^{r}D^{\theta})_{i+1,j}\right).
    \end{split}
\end{equation}
The spatial interpolations are necessary since $H_{r}$ is defined at the same
location as $B^{r}$, but not the same as $B_{\phi}$ (see Figure~\ref{fig:Yee}). A choice can to be made on whether to first multiply the fields by the metric coefficients such as $\alpha\gamma_{r\phi}$ then interpolate, or to first interpolate then multiply by the metric coefficients at the interpolated point. Upon testing, we find that multiplying the coefficients first leads to a more stable field solver, which is adopted in Equation~\eqref{eq:auxH}. This is the algorithm we use in \aperture{}.

Similarly, the update equations for $B^{i}$ are:
\begin{equation}
    \label{eq:B-update}
    \begin{split}
      \frac{\Delta B^{r}_{ij}}{\Delta t} &= -\frac{\left(E_{\phi,i,j+1} - E_{\phi,i,j}\right)\Delta\phi}{A^{B}_{r,ij}} = C_{E}^{r} \\
  \frac{\Delta B^{\theta}_{ij}}{\Delta t} &= -\frac{\left(E_{\phi,i,j} - E_{\phi,i+1,j}\right)\Delta\phi}{A^{B}_{\theta,ij}} = C_{E}^{\theta} \\
  \frac{\Delta B^{\phi}_{ij}}{\Delta t} &= -\frac{\left(E_{\theta,i+1,j} - E_{\theta,i,j}\right)\Delta\theta + \left(E_{r,i,j} - E_{r,i,j+1}\right)\Delta r}{A^{B}_{\phi,ij}} = C_{E}^{\phi}.
    \end{split}
\end{equation}
The auxiliary field components $E_{i}$ are computed using:
\begin{equation}
    \label{eq:auxE}
    \begin{split}
     E_{r,ij} &= \alpha\gamma_{rr}D^{r}_{ij} + \frac{1}{2}\left((\alpha\gamma_{r\phi}D^{\phi})_{ij} + (\alpha\gamma_{r\phi}D^{\phi})_{i+1,j}\right) \\
  E_{\theta,ij} &= \alpha\gamma_{\theta\theta}D^{\theta}_{ij} - \frac{1}{2}\left((\sqrt{\gamma}\beta^{r}B^{\phi})_{ij} + (\sqrt{\gamma}\beta^{r}B^{\phi})_{i-1,j}\right) \\
  E_{\phi,ij} &= \alpha\gamma_{\phi\phi}D^{\phi}_{ij} + \frac{1}{2}\left((\alpha\gamma_{r\phi}D^{r})_{ij} + (\alpha\gamma_{r\phi}D^{r})_{i-1,j}\right) \\
      &\quad + \left((\sqrt{\gamma}\beta^{r}B^{\theta})_{ij} + (\sqrt{\gamma}\beta^{r}B^{\theta})_{i-1,j}\right).
    \end{split}
\end{equation}

\vspace{.1in}

\section{Analytic Initial Conditions}
\label{app:monopole}

The vacuum electromagnetic field solution in the presence of a rotating black hole that is stationary and consistent with a uniform magnetic field of strength $B_{0}$ at large distances was first given by \citet{1974PhRvD..10.1680W}. The solution with a non-rotating black hole ($a = 0$) is given in the main text as Equation~\eqref{eq:vacuum-nonrotating-wald}. The rotating version with nonzero $a$ is used as an initial condition for the plasma-filled problem in Section~\ref{sec:plasma-wald}. The 4-vector potential is given in Boyer-Lindquist coordinates by:
\begin{equation}
  \label{eq:wald-vector-potential}
  A_{\mu}^{BL} = B_{0} \left( \frac{a r(1 + \cos^{2}\theta)}{r^2 + a^2\cos^2\theta} - a, 0, 0, \frac{1}{2}\sin^2\theta \left( r^2 + a^2 - \frac{2a^2 r (1 + \cos^{2}\theta)}{r^2 + a^2\cos^{2}\theta} \right) \right).
\end{equation}
Upon a transformation to the Kerr-Schild coordinates, the $t$ and $\phi$ components of the vector potential remain unchanged, but the $r$ component becomes nonzero:
\begin{equation}
  \label{eq:wald-ks-Ar}
  A_{r}^{KS} = \frac{a^3 (1 - \cos 4 \theta ) - 4a r (r+2) \cos 2 \theta + 4 a(r-6) r}{8 \left(a^2 \cos 2 \theta + a^2 + 2 r^2\right)}.
\end{equation}
In the 3+1 formalism, we define $\Phi=-A_0$, $A^{j}=\gamma^{jk}A_k$, then we have $\bm{E}=-\nabla\Phi-\partial_t \bm{A}$, $\bm{B}=\nabla\times\bm{A}$. The field components in Kerr-Schild coordinates can be computed as:
\begin{align}
  \label{eq:wald-solution-ks}
    B^r&=\frac{1}{\sqrt{\gamma}}\frac{\partial A_{\phi}}{\partial \theta}=B_0\frac{1}{\sqrt{\gamma}}\left( \Delta + \frac{2r(r^4 - a^4)}{\rho^{4}} \right)\sin\theta\cos\theta,\\
    B^\theta&=-\frac{1}{\sqrt{\gamma}}\frac{\partial A_{\phi}}{\partial r}=-B_0\frac{1}{\sqrt{\gamma}}\left( r + a^2(1 + \cos^2\theta) \frac{2r^2 - \rho^2}{\rho^4} \right)\sin^{2}\theta,\\
    B^{\phi}&=-\frac{1}{\sqrt{\gamma}}\frac{\partial A_r}{\partial \theta}=-B_0\frac{1}{\sqrt{\gamma}}\left( \frac{2ar(a^2 - r^2)}{\rho^{4}} - a \right)\sin\theta\cos\theta,\\
    E_r&=-\frac{\partial \Phi}{\partial r}=B_0 a(1 + \cos^2\theta)\frac{a^2\cos^2\theta - r^2}{\rho^{4}},\\
    E_{\theta}&=-\frac{\partial \Phi}{\partial \theta}=B_0\frac{2ar(a^2-r^2)}{\rho^{4}}\sin\theta\cos\theta,\\
    E_{\phi}&=0.
\end{align}
This is also the same electromagnetic field used in the charged particle orbit tests in Section~\ref{sec:trajectories}.

For the monopole initial condition in Section~\ref{sec:application}, we use the following 4-vector potential:
\begin{equation}
    A_{\mu,\text{monopole}}^{KS}=B_0\left(\frac{a \cos\theta}{\rho^2},\frac{a \cos\theta}{\rho^2},0,-\cos\theta\,\frac{r^2+a^2}{\rho^2}\right).
\end{equation}
This is derived from the Boyer-Lindquist coordinate vector potential (given by e.g.\ Crinquand (2021)) using a coordinate transformation. The field components are calculated as follows in the Kerr-Schild coordinates
\begin{align}
    B^r&=\frac{1}{\sqrt{\gamma}}\frac{\partial A_{\phi}}{\partial \theta}=B_0\frac{1}{\sqrt{\gamma}}\frac{(a^2+r^2)(r^2-a^2\cos^2\theta)\sin\theta}{\rho^4},\\
    B^\theta&=-\frac{1}{\sqrt{\gamma}}\frac{\partial A_{\phi}}{\partial r}=-B_0\frac{1}{\sqrt{\gamma}}\frac{2a^2r\sin^2\theta\cos\theta}{\rho^4},\\
    B^{\phi}&=-\frac{1}{\sqrt{\gamma}}\frac{\partial A_r}{\partial \theta}=B_0\frac{1}{\sqrt{\gamma}}\frac{a(r^2-a^2\cos^2\theta)\sin\theta}{\rho^4},\\
    E_r&=-\frac{\partial \Phi}{\partial r}=-B_0\frac{2ar\cos\theta}{\rho^4},\\
    E_{\theta}&=-\frac{\partial \Phi}{\partial \theta}=-B_0\frac{a(r^2-a^2\cos^2\theta)\sin\theta}{\rho^4},\\
    E_{\phi}&=0.
\end{align}
This is the initial field configuration we use in our simulations.


\begin{thebibliography}{}
\expandafter\ifx\csname natexlab\endcsname\relax\def\natexlab#1{#1}\fi
\providecommand{\url}[1]{\href{#1}{#1}}
\providecommand{\dodoi}[1]{doi:~\href{http://doi.org/#1}{\nolinkurl{#1}}}
\providecommand{\doeprint}[1]{\href{http://ascl.net/#1}{\nolinkurl{http://ascl.net/#1}}}
\providecommand{\doarXiv}[1]{\href{https://arxiv.org/abs/#1}{\nolinkurl{https://arxiv.org/abs/#1}}}

\bibitem[{Alexandrescu(2001)}]{10.5555/377789}
Alexandrescu, A. 2001, Modern C++ design: generic programming and design
  patterns applied (USA: Addison-Wesley Longman Publishing Co., Inc.)

\bibitem[{{Bacchini} {et~al.}(2018){Bacchini}, {Ripperda}, {Chen}, \&
  {Sironi}}]{2018ApJS..237....6B}
{Bacchini}, F., {Ripperda}, B., {Chen}, A.~Y., \& {Sironi}, L. 2018, \apjs,
  237, 6, \dodoi{10.3847/1538-4365/aac9ca}

\bibitem[{{Bacchini} {et~al.}(2019){Bacchini}, {Ripperda}, {Porth}, \&
  {Sironi}}]{2019ApJS..240...40B}
{Bacchini}, F., {Ripperda}, B., {Porth}, O., \& {Sironi}, L. 2019, \apjs, 240,
  40, \dodoi{10.3847/1538-4365/aafcb3}

\bibitem[{{Beloborodov}(2008)}]{2008ApJ...683L..41B}
{Beloborodov}, A.~M. 2008, \apjl, 683, L41, \dodoi{10.1086/590079}

\bibitem[{{Bhattacharjee} {et~al.}(2009){Bhattacharjee}, {Huang}, {Yang}, \&
  {Rogers}}]{2009PhPl...16k2102B}
{Bhattacharjee}, A., {Huang}, Y.-M., {Yang}, H., \& {Rogers}, B. 2009, Physics
  of Plasmas, 16, 112102, \dodoi{10.1063/1.3264103}

\bibitem[{{Birdsall} \& {Langdon}(1991)}]{1991ppcs.book.....B}
{Birdsall}, C.~K., \& {Langdon}, A.~B. 1991, {Plasma Physics via Computer
  Simulation}

\bibitem[{{Blandford} \& {Znajek}(1977)}]{1977MNRAS.179..433B}
{Blandford}, R.~D., \& {Znajek}, R.~L. 1977, \mnras, 179, 433,
  \dodoi{10.1093/mnras/179.3.433}

\bibitem[{{Boris}(1970)}]{boris_relativistic_1970}
{Boris}, J.~P. 1970, Proceeding of Fourth Conference on Numerical Simulations
  of Plasmas

\bibitem[{{Bransgrove} {et~al.}(2021){Bransgrove}, {Ripperda}, \&
  {Philippov}}]{2021PhRvL.127e5101B}
{Bransgrove}, A., {Ripperda}, B., \& {Philippov}, A. 2021, \prl, 127, 055101,
  \dodoi{10.1103/PhysRevLett.127.055101}

\bibitem[{{Broderick} \& {Tchekhovskoy}(2015)}]{2015ApJ...809...97B}
{Broderick}, A.~E., \& {Tchekhovskoy}, A. 2015, \apj, 809, 97,
  \dodoi{10.1088/0004-637X/809/1/97}

\bibitem[{{Chen}(2017)}]{2017PhDT.......278C}
{Chen}, A. 2017, PhD thesis, Columbia University, New York

\bibitem[{{Chen} \& {Beloborodov}(2014)}]{2014ApJ...795L..22C}
{Chen}, A.~Y., \& {Beloborodov}, A.~M. 2014, \apjl, 795, L22,
  \dodoi{10.1088/2041-8205/795/1/L22}

\bibitem[{{Chen} \& {Yuan}(2020)}]{2020ApJ...895..121C}
{Chen}, A.~Y., \& {Yuan}, Y. 2020, \apj, 895, 121,
  \dodoi{10.3847/1538-4357/ab8c46}

\bibitem[{{Cheng} {et~al.}(1986){Cheng}, {Ho}, \&
  {Ruderman}}]{1986ApJ...300..500C}
{Cheng}, K.~S., {Ho}, C., \& {Ruderman}, M. 1986, \apj, 300, 500,
  \dodoi{10.1086/163829}

\bibitem[{{Crinquand} {et~al.}(2021){Crinquand}, {Cerutti}, {Dubus}, {Parfrey},
  \& {Philippov}}]{2021A&A...650A.163C}
{Crinquand}, B., {Cerutti}, B., {Dubus}, G., {Parfrey}, K., \& {Philippov}, A.
  2021, \aap, 650, A163, \dodoi{10.1051/0004-6361/202040158}

\bibitem[{{Crinquand} {et~al.}(2022){Crinquand}, {Cerutti}, {Dubus}, {Parfrey},
  \& {Philippov}}]{2022PhRvL.129t5101C}
---. 2022, \prl, 129, 205101, \dodoi{10.1103/PhysRevLett.129.205101}

\bibitem[{{Crinquand} {et~al.}(2020){Crinquand}, {Cerutti}, {Philippov},
  {Parfrey}, \& {Dubus}}]{2020PhRvL.124n5101C}
{Crinquand}, B., {Cerutti}, B., {Philippov}, A., {Parfrey}, K., \& {Dubus}, G.
  2020, \prl, 124, 145101, \dodoi{10.1103/PhysRevLett.124.145101}

\bibitem[{{El Mellah} {et~al.}(2023){El Mellah}, {Cerutti}, \&
  {Crinquand}}]{2023A&A...677A..67E}
{El Mellah}, I., {Cerutti}, B., \& {Crinquand}, B. 2023, \aap, 677, A67,
  \dodoi{10.1051/0004-6361/202346781}

\bibitem[{{El Mellah} {et~al.}(2022){El Mellah}, {Cerutti}, {Crinquand}, \&
  {Parfrey}}]{2022A&A...663A.169E}
{El Mellah}, I., {Cerutti}, B., {Crinquand}, B., \& {Parfrey}, K. 2022, \aap,
  663, A169, \dodoi{10.1051/0004-6361/202142847}

\bibitem[{{Esirkepov}(2001)}]{2001CoPhC.135..144E}
{Esirkepov}, T.~Z. 2001, Computer Physics Communications, 135, 144,
  \dodoi{10.1016/S0010-4655(00)00228-9}

\bibitem[{{Galishnikova} {et~al.}(2023){Galishnikova}, {Philippov}, {Quataert},
  {Bacchini}, {Parfrey}, \& {Ripperda}}]{2023PhRvL.130k5201G}
{Galishnikova}, A., {Philippov}, A., {Quataert}, E., {et~al.} 2023, \prl, 130,
  115201, \dodoi{10.1103/PhysRevLett.130.115201}

\bibitem[{{Garofalo} \& {Meier}(2010)}]{2010MNRAS.406.2047G}
{Garofalo}, D., \& {Meier}, D.~L. 2010, \mnras, 406, 2047,
  \dodoi{10.1111/j.1365-2966.2010.16815.x}

\bibitem[{{Guo} {et~al.}(2014){Guo}, {Li}, {Daughton}, \&
  {Liu}}]{2014PhRvL.113o5005G}
{Guo}, F., {Li}, H., {Daughton}, W., \& {Liu}, Y.-H. 2014, \prl, 113, 155005,
  \dodoi{10.1103/PhysRevLett.113.155005}

\bibitem[{{Haugb{\o}lle} {et~al.}(2013){Haugb{\o}lle}, {Frederiksen}, \&
  {Nordlund}}]{2013PhPl...20f2904H}
{Haugb{\o}lle}, T., {Frederiksen}, J.~T., \& {Nordlund}, A. 2013, Physics of
  Plasmas, 20, 062904, \dodoi{10.1063/1.4811384}

\bibitem[{{Kin} {et~al.}(2024){Kin}, {Kisaka}, {Toma}, {Kimura}, \&
  {Levinson}}]{2024ApJ...964...78K}
{Kin}, K., {Kisaka}, S., {Toma}, K., {Kimura}, S.~S., \& {Levinson}, A. 2024,
  \apj, 964, 78, \dodoi{10.3847/1538-4357/ad20cd}

\bibitem[{King {et~al.}(1975)King, Lasota, \& Kundt}]{PhysRevD.12.3037}
King, A.~R., Lasota, J.~P., \& Kundt, W. 1975, Phys. Rev. D, 12, 3037,
  \dodoi{10.1103/PhysRevD.12.3037}

\bibitem[{{Kisaka} {et~al.}(2020){Kisaka}, {Levinson}, \&
  {Toma}}]{2020ApJ...902...80K}
{Kisaka}, S., {Levinson}, A., \& {Toma}, K. 2020, \apj, 902, 80,
  \dodoi{10.3847/1538-4357/abb46c}

\bibitem[{{Komissarov}(2004)}]{2004MNRAS.350..427K}
{Komissarov}, S.~S. 2004, \mnras, 350, 427,
  \dodoi{10.1111/j.1365-2966.2004.07598.x}

\bibitem[{{Kunz} {et~al.}(2014){Kunz}, {Schekochihin}, \&
  {Stone}}]{2014PhRvL.112t5003K}
{Kunz}, M.~W., {Schekochihin}, A.~A., \& {Stone}, J.~M. 2014, \prl, 112,
  205003, \dodoi{10.1103/PhysRevLett.112.205003}

\bibitem[{{Levin} \& {Perez-Giz}(2008)}]{2008PhRvD..77j3005L}
{Levin}, J., \& {Perez-Giz}, G. 2008, \prd, 77, 103005,
  \dodoi{10.1103/PhysRevD.77.103005}

\bibitem[{{Levinson} \& {Cerutti}(2018)}]{2018A&A...616A.184L}
{Levinson}, A., \& {Cerutti}, B. 2018, \aap, 616, A184,
  \dodoi{10.1051/0004-6361/201832915}

\bibitem[{{McKinney} {et~al.}(2012){McKinney}, {Tchekhovskoy}, \&
  {Blandford}}]{2012MNRAS.423.3083M}
{McKinney}, J.~C., {Tchekhovskoy}, A., \& {Blandford}, R.~D. 2012, \mnras, 423,
  3083, \dodoi{10.1111/j.1365-2966.2012.21074.x}

\bibitem[{{Niv} {et~al.}(2023){Niv}, {Bromberg}, {Levinson}, {Cerutti}, \&
  {Crinquand}}]{2023MNRAS.526.2709N}
{Niv}, I., {Bromberg}, O., {Levinson}, A., {Cerutti}, B., \& {Crinquand}, B.
  2023, \mnras, 526, 2709, \dodoi{10.1093/mnras/stad2904}

\bibitem[{NVIDIA {et~al.}(2024)NVIDIA, Vingelmann, \& Fitzek}]{cuda}
NVIDIA, Vingelmann, P., \& Fitzek, F.~H. 2024, CUDA, release: 12.3.2.
\newblock \url{https://developer.nvidia.com/cuda-toolkit}

\bibitem[{{Parfrey} {et~al.}(2019){Parfrey}, {Philippov}, \&
  {Cerutti}}]{2019PhRvL.122c5101P}
{Parfrey}, K., {Philippov}, A., \& {Cerutti}, B. 2019, \prl, 122, 035101,
  \dodoi{10.1103/PhysRevLett.122.035101}

\bibitem[{{Qin} {et~al.}(2013){Qin}, {Zhang}, {Xiao}, {Liu}, {Sun}, \&
  {Tang}}]{2013PhPl...20h4503Q}
{Qin}, H., {Zhang}, S., {Xiao}, J., {et~al.} 2013, Physics of Plasmas, 20,
  084503, \dodoi{10.1063/1.4818428}

\bibitem[{{Sironi} \& {Spitkovsky}(2014)}]{2014ApJ...783L..21S}
{Sironi}, L., \& {Spitkovsky}, A. 2014, \apjl, 783, L21,
  \dodoi{10.1088/2041-8205/783/1/L21}

\bibitem[{{Tchekhovskoy} {et~al.}(2011){Tchekhovskoy}, {Narayan}, \&
  {McKinney}}]{2011MNRAS.418L..79T}
{Tchekhovskoy}, A., {Narayan}, R., \& {McKinney}, J.~C. 2011, \mnras, 418, L79,
  \dodoi{10.1111/j.1745-3933.2011.01147.x}

\bibitem[{{Umeda} {et~al.}(2003){Umeda}, {Omura}, {Tominaga}, \&
  {Matsumoto}}]{2003CoPhC.156...73U}
{Umeda}, T., {Omura}, Y., {Tominaga}, T., \& {Matsumoto}, H. 2003, Computer
  Physics Communications, 156, 73, \dodoi{10.1016/S0010-4655(03)00437-5}

\bibitem[{{Vay}(2008)}]{2008PhPl...15e6701V}
{Vay}, J.~L. 2008, Physics of Plasmas, 15, 056701, \dodoi{10.1063/1.2837054}

\bibitem[{{Villasenor} \& {Buneman}(1992)}]{1992CoPhC..69..306V}
{Villasenor}, J., \& {Buneman}, O. 1992, Computer Physics Communications, 69,
  306, \dodoi{10.1016/0010-4655(92)90169-Y}

\bibitem[{{Wald}(1974)}]{1974PhRvD..10.1680W}
{Wald}, R.~M. 1974, \prd, 10, 1680, \dodoi{10.1103/PhysRevD.10.1680}

\bibitem[{{Werner} {et~al.}(2016){Werner}, {Uzdensky}, {Cerutti}, {Nalewajko},
  \& {Begelman}}]{2016ApJ...816L...8W}
{Werner}, G.~R., {Uzdensky}, D.~A., {Cerutti}, B., {Nalewajko}, K., \&
  {Begelman}, M.~C. 2016, \apjl, 816, L8, \dodoi{10.3847/2041-8205/816/1/L8}

\bibitem[{{Yee}(1966)}]{1966ITAP...14..302Y}
{Yee}, K. 1966, IEEE Transactions on Antennas and Propagation, 14, 302,
  \dodoi{10.1109/TAP.1966.1138693}

\bibitem[{{Yuan} \& {Narayan}(2014)}]{2014ARA&A..52..529Y}
{Yuan}, F., \& {Narayan}, R. 2014, \araa, 52, 529,
  \dodoi{10.1146/annurev-astro-082812-141003}

\end{thebibliography}

\end{document}